\documentclass[9pt,twocolumn,twoside]{opticajnl}
\journal{opticajournal} 

\usepackage{amsmath,amsfonts,amssymb}
\usepackage{graphicx}
\usepackage[colorlinks=true, allcolors=blue]{hyperref}
\usepackage{float}
\usepackage{esvect}
\usepackage[export]{adjustbox}
\usepackage{subcaption}
\usepackage{hyperref}
\usepackage[normalem]{ulem} 
\newcommand{\unit}[1]{\,\mathrm{#1}}
\setboolean{shortarticle}{true}

\usepackage[colorlinks=true, allcolors=blue]{hyperref}
\usepackage{lineno}
\usepackage{tabularx}

\title{Phasing segmented telescopes via deep learning methods: application to a deployable CubeSat}

\author[1,2,3,*]{ Maxime DUMONT}
\author[2,4]{Carlos M. CORREIA}
\author[1,3]{Jean-François SAUVAGE}
\author[5]{Noah SCHWARTZ}
\author[3]{Morgan GRAY}
\author[2,6]{Jaime CARDOSO}

\affil[1]{DOTA, ONERA, F-13661 Salon Cedex Air - France}
\affil[2]{Faculdade de Engenharia da Universidade do Porto, Rua Dr. Roberto Frias, s/n, 4200-465 Porto, Portugal}
\affil[3]{Aix-Marseille Université, CNRS, CNES, LAM, Marseille, France}
\affil[4]{Center for Astrophysics and Gravitation, Instituto Superior Técnico, Av. Rovisco Pais 1, 1049-001 Lisboa, Portugal}
\affil[5]{UK Astronomy Technology Centre, Edinburgh EH9 3HJ, United Kingdom}
\affil[6]{INESCTEC, Porto, Portugal}
\affil[*]{maxime.dumont@lam.fr}

\begin{abstract}

Capturing high-resolution imagery of the Earth's surface often calls for a telescope of considerable size, even from Low Earth Orbits (LEO). A large aperture often requires large and expensive platforms. For instance, achieving a resolution of 1\,m at visible wavelengths from LEO typically requires an aperture diameter of at least 30\,cm. Additionally, ensuring high revisit times often prompts the use of multiple satellites. 
In light of these challenges, a small, segmented, deployable CubeSat telescope was recently proposed creating the additional need of phasing the telescope’s mirrors. Phasing methods on compact platforms are constrained by the limited volume and power available, excluding solutions that rely on dedicated hardware or demand substantial computational resources. 
Neural Network (NN) are known for their computationally efficient inference and reduced on-board requirements. Therefore we developed a NN-based method to measure co-phasing errors inherent to a deployable telescope.
The proposed technique demonstrates its ability to detect phasing error at the targeted performance level (typically a wavefront error (WFE) below 15\,nm\,RMS for a visible imager operating at the diffraction limit) using a point source. The robustness of the NN method is verified in presence of high-order aberrations or noise and the results are compared against existing state-of-the-art techniques. 
The developed NN model ensures its feasibility and provides a realistic pathway towards achieving diffraction-limited images.

\end{abstract}

\begin{document} 
\maketitle


\section{Context and main challenges}


The angular resolution of telescopes is fundamentally limited by the size of their aperture for a given wavelength. For example, a telescope with a diameter of $30\unit{cm}$ has a ground sampling of $1\unit{m}$ at a distance of $400\unit{km}$ or a $2\unit{m}$ at $800\unit{km}$ at visible wavelengths ($\lambda = 800\unit{nm}$). In the fields of Earth observation and astronomy, high-angular resolution images are crucial for maximizing the scientific output and return on investment, which has driven the demand for larger apertures. However, launching large monolithic pupils into space is expensive, and designing and manufacturing such telescopes pose significant challenges$\,$\cite{sabelhaus2004overview}. An effective solution to increase aperture size is to fragment the aperture into smaller segments. By folding the telescope inside a compact volume during launch and deploying it in orbit, both the collecting power and angular resolution can be increased.  This volume optimization and gain in mass translates directly into cost savings, enabling the deployment of a constellation with multiple high-angular resolution platforms and therefore achieving higher revisit rate (i.e. temporal resolution).

Improving both the revisit rate and angular resolution finds applications in Earth climate monitoring, defence, and security for which the highest commercially-available spatial resolutions today are given by the satellite Pleiades (with a Ground Sampling Distance (GSD) of 70\,cm, at a rate of 1 image per day) as well as Pleiades-Neo (30\,cm GSD colour images). For example, flood monitoring  (accounting for over 40\% of the natural disasters)  currently provides information either over large areas or with limited spatial resolutions and a revisit every two days (e.g. Copernicus with a 5\,km resolution), or instead with a revisit time at best of a few days yet with a high spatial resolution (e.g. Pléiades, Sentinel-1, and Sentinel-2). Solar system exploration can potentially also benefit from this improvement: although the HRI camera of Deep Impact has an aperture larger than 30\,cm, the typical size is 10 to 20\,cm (NAC camera on Rosetta, Cassis camera on TGO, New Horizon camera). 

\begin{figure}[h!]
\begin{center}
\includegraphics[height=6cm]{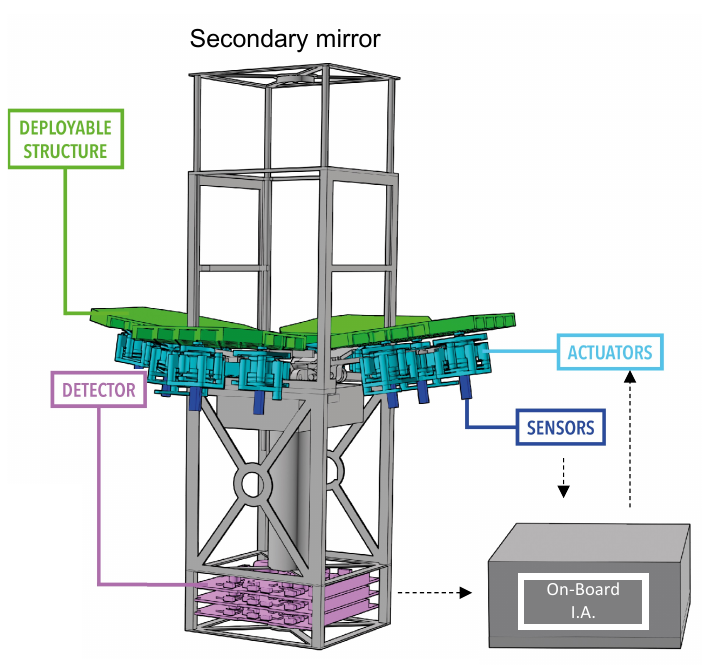}
\caption[example] 
{   \label{fig:Nanosat} 
    Simplified illustration of the deployable payload concept: deployable structures (primary \& secondary mirrors, baffle), actuators to adjust mirror positions, sensors to measure mirror positions, detector to assess image quality, and onboard artificial intelligence to control and adjust mirror positions to reach diffraction-limited image quality (i.e. active optics). From N. Schwartz$\,$\cite{Schwartz}.}
\end{center}
\end{figure}

The objective of this study is to enable, in \href{https://business.esa.int/newcomers-earth-observation-guide#ref_2.1.1}{ESA's} own parlance \textit{very very high-angular resolution} imaging in the visible spectrum for Earth Observation from space at a low cost by quantifying the performance of a NN-based phasing algorithm during pre-launch, deployment and exploitation phases. To achieve this, we use AZIMOV$\,$\cite{Schwartz} as a reference design for our study. AZIMOV is a satellite concept featuring a $30\unit{cm}$ wide, diluted aperture telescope onboard a 6U CubeSat. It is capable of reaching a $1\unit{m}$ GSD in the visible (i.e. $GSD = \lambda/D *z=800.10^{-9}/0.3\times 400000 \approx 1\unit{m}$) from LEO with a swath of over 4\,km. Fig.\,\ref{fig:Nanosat} illustrates a simplified concept of the payload. This technology allows the telescope's aperture to surpass the size of the supporting platform, thereby enhancing the telescope's light-gathering power and improving its achievable resolution.

To achieve the theoretical angular resolution for a given segmented pupil, the individual mirror segments must be co-phased with a precision typically on the order of a fraction of the imaging wavelength (usually $\lambda/14$$\,$\cite{born2013principles}). Compensation for static and dynamic phase aberrations is required for each segment. High-precision actuators are utilized to control the position of the deployable primary mirror segments in Piston and Tip-Tilt$\,$\cite{schwartz20226u} (PTT). 
For visible wavelength imaging at the metric resolution limit (we select $\lambda = 800\unit{nm}$ as the study wavelength), the WaveFront Error (WFE) after correction should be less than $50\unit{nm}\ RMS$. However, due to actuator response and dynamics, a portion of the WFE error is independent of sensing. From a dedicated study of this error budget, Sauvage\,\cite{sauvage2020first} allocate  $\epsilon_{WFS} < 15\unit{nm\ RMS}$ to the sensing and reconstruction contribution. 
Our study proposes a solution to perform accurate WFS onboard a CubeSat, using directly a single focal plane image and Deep Learning.

This paper is organized as follows: we first provide an introduction to the study and the motivation behind it. Secondly we delve into the  concepts of image formation in the context of wavefront sensing. The methodology employed in our research is detailed in Part.\,\ref{sec:Methods}. Then, Part.\,\ref{sec:Res} presents the results obtained from our experiments, analyzing the performance and robustness of the proposed methods. Part.\,\ref{sec:AZI} discusses the implementation of neural networks in a complete phasing process specifically tailored for the AZIMOV system. Finally, Part.\,\ref{sec:Ccl} concludes the paper.

\section{Phasing segmented mirror telescopes}

The phasing of segmented mirror telescopes is a critical aspect of their operation, as it directly impacts the quality of their optical performance.
Ground-based telescopes such as the segmented, twin Keck telescopes correct for tip-tilt errors using a Shack-Hartmann WaveFront Sensor (SH-WFS)\, \cite{chanan1986segment}. However, the SH-WFS is known to be insensitive to global piston error. To overcome this issue, a micro-lenslet array is placed at the junction of two segments on the primary mirror, and allows to identify the deferential piston between 2 segments$\,$\cite{chanan1986segment}. More recently, for the Giant Magellan Telescope, and following roughly the same principle, differential piston between segments will be identified using a Holographic Dispersed Fringe Sensor (HDFS)$\,$\cite{haffert2022phasing}. It employs a holographic optical element to disperse the incoming light into multiple wavelength channels. By analyzing the interference patterns created by these channels, the HDFS can accurately measure the differential piston.
For space-borne telescopes such as the JWST\,\cite{acton2022phasing}, the phasing strategy is divided in 3 parts: i) the Segment Location-Identification, to identify the coarse position of each segment ii) the Co-Phasing to correct the global phasing error of the 18 segments. iii) finally the Wavefront Monitoring and Maintenance to ensure a correct phasing throughout the life of the mission. The coarse and fine phasing is ensured by a dispersed fringe SH-WFS and weak lens hardware$\,$\cite{perrin2016preparing}.
In this context, this step-by-step approach has inspired our work. However, we propose an innovative approach to co-phase segmented telescopes by using a single focal plane image as the observable, and deep learning to tune our NN-based algorithm. Single-image approaches can greatly simplify the phasing process and enhance the optical performance of future segmented telescopes. Our work focuses specifically on the primary mirror phasing from the Point Spread Functions (PSF) obtained while imaging a point source object. The use of a point source is a key element in our approach. First to understand the behaviour and NN specifics to process focal-plane images and estimate wavefront errors from them. Secondly, to be able to generalise to extended objects. Lastly, the phasing process of AZIMOV isn't defined yet. One strategy could be to undertake the coarse phasing imaging a distant unresolved star and then rotate the telescope towards the Earth to proceed with the fine phasing and capture images for further ends. The use of NN on point sources is therefore central to many applications. Our work aims to employ a computationally efficient phasing method taking into account the typical CubeSat limitations especially in terms of volume and computing power. In particular, we demonstrate the potential of using NN in contrast to more classical optimization methods, focusing on execution time as a proxy to computational complexity. In this context, we use the time required for NN inference to identify a suitable NN model capable of reaching the diffraction limit with minimal computational burden. The AZIMOV payload is currently being designed, and requirements for the on-board computer (OBC) are still being finalized. Among other aspects, the OBC will need to run the compute intensive co-phasing algorithm, the data-intensive image processing algorithm (including algorithms such as image registration and super-resolution) and other housekeeping tasks. The satellite being in LEO, it will receive a varying thermal load (approx. every 90 min): multiple updates per orbit will be required to maintain good phasing performance. It is important to minimize the cophasing algorithm complexity and therefore execution time. The exact execution time will naturally depend on the selected OBC.

\subsection{Classical wavefront sensing approaches}
Various methods have been proposed for measuring and estimating wavefront error, broadly classified into two categories: direct wavefront sensing and focal plane methods. Direct wavefront sensing relies on relatively simple algorithm, at a cost of an optical path specific to the wavefront sensing with dedicated hardware, such as the JWST's dispersed fringe Shack-Hartmann WFS\,\cite{acton2022phasing}.

An alternative method involves indirect measurement, where the wavefront is deduced from the focal-plane image captured by the detector. One approach is image sharpening, as proposed in the work of Schwartz$\,$\cite{Schwartz,2017JATIS...3c9001L} which entails acquiring a sequence of images and iterate starting from an initial guess of the aberrations. For each of these images, an image quality metric, such as maximum intensity or image contrast, is computed and then used in an non-linear iterative optimization algorithm to estimate the aberration. 
In contrast, model-based focal plane wavefront sensing methods typically rely on introducing diversity (whether in phase e.g Phase Diversity \cite{mugnier2006phase}, wavelength, or amplitude) to resolve sign-ambiguity \cite{martinache2013asymmetric} of even modes. A best fit is sought to the focal plane images where from the wavefront aberration parameters are estimated. The algorithms are typically more complex than those used in direct wavefront sensing due to the relationship between the image intensity in the focal plane and the electric field in the pupil plane. Nevertheless, considering the constraints posed by Earth Observation and the non-repeatability of scenes, the goal is to conduct focal-plane wavefront sensing using only a single image.
Data-driven, Deep learning based methods can also be applied successfully in Focal Plane WaveFront Sensing (FPWFS). One of the main advantages lies in the fact that once the network is trained, the model directly outputs the estimated wavefront without requiring iterations. The computation gains time and, consequently, efficiency in calculations. The network's parameters have been learned during the training phase, where all the computational burden is handled before launch. Therefore, once on-board, the process is relatively fast and not contingent on the uncertainty of the number of iterations. The training steps build numerically and iteratively the NN model from the data. Tab.\ref{tab:methods} compares qualitatively the 3 methods cited above.

\begin{table}
\begin{center}
\resizebox{0.45\textwidth}{!}{
\begin{tabular} {||c || c c c||}
 \hline
   & Deep learning method & Image sharpening & Phase Diversity\\ 
 \hline\hline
 Data Driven & Yes & No & No\\ 
 \hline
 Model-free& Yes & Yes & No  \\
 \hline
 Iterative & No & Very & Yes \\
 \hline
 On-board computing & Deterministic & Optical optimization &  Numerical optimization \\
 \hline
 Off-board computing & Stochastic & None &  None \\
 \hline
 Initial guess & No & Yes & Yes \\
 \hline
\end{tabular}}
\caption{Feature comparison of FPWFS methods.}
\label{tab:methods}
\end{center}
\end{table}

\subsection{Motivation for new phasing approaches for  space-borne segmented telescopes}

Recently, access to space has shown significant changes characterized by the emergence of the 'NewSpace' paradigm, which emphasizes cost-effective access to space and introduces novel constraints such as limited volume allocation. Therefore, this necessitates the development of innovative methodologies, such as those shown in this paper.
By leveraging a single focal plane image, we estimate and subsequently correct low-order phase aberrations piston and tip-tilt using Neural Networks (NN) and a space-borne, 4-petal deployable CubeSat as reference design. CubeSats are not only constrained in volume but also in computing power. Therefore, the complexity of the model will be considered in this study, as it plays a crucial role in minimizing optical aberrations control for successful implementation on small satellites. 

\subsection{Previous work on NN for focal-plane wavefront sensing}
The utilisation of deep learning techniques has been proposed in previous studies as a powerful tool for FPWFS, often aiming for performance levels similar or beyond the state-of-the-art\,\cite{rossi2022machine, herbel2018fast,paine2018machine}. The majority of NN methods for FPWFS employs a single defocused images, as demonstrated in studies such as \cite{rossi2022machine, wang2021deep, rajaoberison2022machine, andersen2020image, paine2018machine, orban2021focal}. Others have used an optical preconditioner (overexposure, defocus, scatter) to improve the performance of the WFS $\,$\cite{Nishizaki}. Finally, others have used a pair of in and out of focus images to identify Zernike coefficients from a pair of PSF. $\,$\cite{quesnel2022deep}.
These works have shown excellent performance for WFSing using deep NN architectures, with Xception and Inception v3 being the most widely used. Xception utilises a depthwise separable convolution approach, which separates spatial and channel-wise convolution operations. These architectures have been for example employed by Paine$\,$\cite{paine2018machine}, Rajaoberison$\,$\cite{rajaoberison2022machine}, Andersen$\,$\cite{andersen2020image}, Nishizaki$\,$\cite{Nishizaki}, and Orban$\,$\cite{orban2021focal}. Other popular architectures such as ResNet 50 and VGG-16, have also been tested by Paine$\,$\cite{paine2018machine} and Orban$\,$\cite{orban2021focal}. Architectures like EfficientNet-B4 (used by Quesnel$\,$\cite{quesnel2022deep}), Bi-GRU for sequential data (employed by Wang$\,$\cite{wang2021deep}), and Dense NN (used by Rossi$\,$\cite{rossi2022machine}) also have demonstrated effectiveness for focal plane wavefront sensing.
In these studies, two types of focal plane wavefront sensing have been identified. The majority of them focusing on Zernike mode identification for a monolithic pupil imager (Paine$\,$\cite{paine2018machine}, Andersen$\,$\cite{andersen2020image}, Nishizaki$\,$\cite{Nishizaki}, Orban$\,$\cite{orban2021focal}, and Quesnel$\,$\cite{quesnel2022deep}), while others aim at phasing a segmented pupil by identifying low-order modes (Wang$\,$\cite{wang2021deep} focuses on Piston and Tip-Tilt, Rossi$\,$\cite{rossi2022machine} and Rajaoberison$\,$\cite{rajaoberison2022machine} focus on Piston only).
The major shortcoming of these studies is that the NN architecture requires hundreds of thousands to hundreds of millions of parameters (e.g., ResNet 50 and VGG-16 require respectively 23 million and 138 million parameters). Our goal, apart from dealing with specific segmentation errors, is to minimise the number of parameters in order to optimize the inference time. For this reason we need to explore new suitable architectures will smaller dimensionalities while keeping a bottleneck architecture like, such as VGG-net and Resnet.

\section{Forward model}\label{sec:IF}

\subsection{Image formation}
The image of a point source, or PSF, can be described as:
    \begin{equation}
    \label{eq:PSF}
     PSF = |\text{FT} (P e^{j\phi})|^{2} = |\text{FT} (P e^{j(\phi_{PTT} + \phi_{HO})})|^{2}   
    \end{equation}
 where FT is the Fourier Transform, $P$ is the pupil transmission function, with 1 inside the pupil boundaries and 0 outside. $\phi$ is the phase function which can be decomposed as $\phi = \phi_{PTT} + \phi_{HO}$  corresponding respectively to the PTT phase and the Higher Order (HO) phase. Also, $\phi=\frac{2\pi \text{$\Delta$}}{\lambda}$ where $\Delta$ the corresponding Optical Path Difference (OPD) and the wavelength $\lambda = 800\unit{nm}$. \\
 
\begin{figure}[h!]
\begin{center}
\includegraphics[height=4cm]{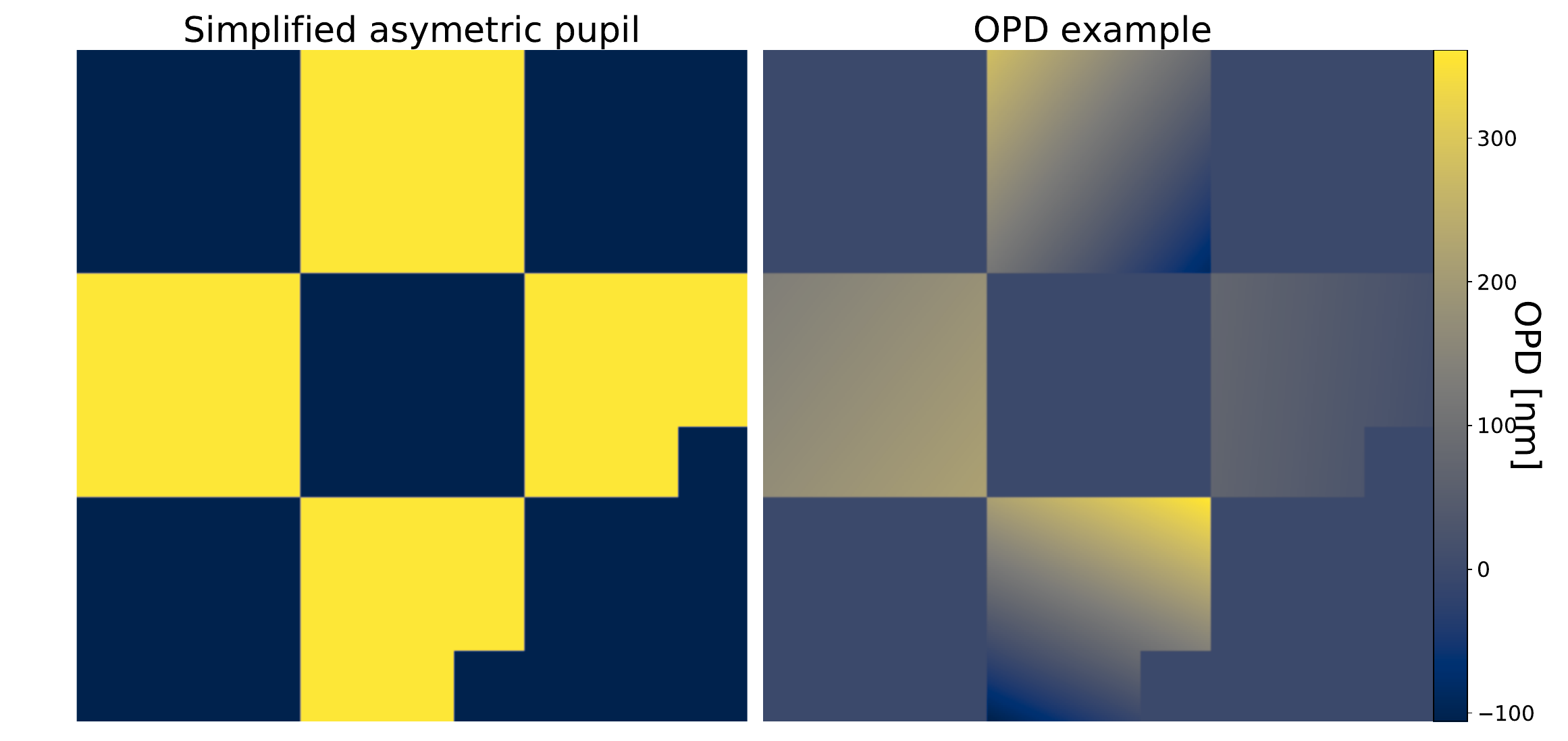}
\caption[example] 
{   \label{fig:pup} 
Example of the pupil and OPD map.}
\end{center}
\end{figure}

The AZIMOV prototype (Fig.\,\ref{fig:Nanosat}) is composed of a pupil with 4 deployable petals. The precise design of AZIMOV pupil is not finalized today. The global shape is four segments with an approximately square shape, but the final design might evolve due to the reduced volume constraints. In particular the four segments might not be symmetric in the final design. In this paper, we propose a simplified pupil composed of 4 squared segments where 2 segments are $10\%$ cropped on the edge in order to break the pupil centro-symmetry as shown in Fig.\,\ref{fig:pup}.  The asymmetric pupil Fourier wavefront sensor (APF-WFS) in Martinache's work shows great potential for small aberrations on an asymmetric pupil. Pope$\,$\cite{pope2014demonstration} shows an interest in using Martimache’s work for segmented pupil, retrieving piston value for an interval of $\frac{\lambda}{3.5}$. However it hasn’t been studied for large tip-tilt errors, a subject discussed in this paper. This method appears to be faster than our approach (only requiring a FFT) but lacks the ability to estimate large aberrations and only works on a point source - essential considerations for AZIMOV.  In our case, cropping 10\% of the segments is chosen as a trade-off between the loss of collecting area and the correct wavefront estimation: the more asymmetric the pupil is, the better the even parts of the phase will be identified\,\cite{martinache2013asymmetric}. However a limit is shown in the ratio of even part identification over the percentage of cropped area. For instance, doubling the cropped area (from 10\% to 20\% of 2 segments) improves the estimation by 25\%. For our space-borne application, the pupil design and the need to use the images from the science detector directly compels us to use instead an amplitude diversity as shown in Fig.\ref{fig:pup}.

To address the specifics of a segmented pupil, we define our modal basis with three modes $B_{i=1..3}^{k=1..4}$ on each of the four segments, i.e.
\begin{equation}
    \phi = \sum_{i=1}^3 \sum_{k=1}^4 c_i^k B_i^k
\end{equation}
where the modes are
\begin{equation}\label{eq:modalBasis}
    \begin{array}{cl}
        B_1^k(x,y) & =  4  \\
        B_2^k(x,y) & = 4\sqrt{3} x \\
        B_3^k(x,y) & = 4\sqrt{3} y
    \end{array}
\end{equation}
representing respectively Piston, Tip and Tilt on segment $k$. Clearly these modes are orthogonal across different segments and their inner product.

\begin{equation}\label{eq:scalarProduct}
    \langle B_i, B_j\rangle = \left\{ 
    \begin{array}{cc}
        1 &  \text{if}\,\,\, i = j\\
        0  & \text{otherwise}
    \end{array}\right.
\end{equation}

The expansion coefficients in Eq. (\ref{eq:modalBasis}) are easily found when establishing an orthonormalised basis with unitary root-mean-square (RMS) over the whole, un-cropped, pupil in Fig.\ref{fig:pup}
\begin{equation}\label{eq:RMS}
    \frac{1}{4}\sqrt{\frac{1}{4}\int_{-1}^1\int_{-1}^1 M_i M_j dx dy} = 1 \,\,\,\,\,\hspace{20pt}\text{if}\,\,\, i=j
\end{equation}
where the $1/4$ in front of the square root represents one-fourth of the pupil, i.e. the relative area of one segment, and the integral is computed on normalised units; and $M_i$ represents the  phase mode of one squared pupil. For instance $M_1$ corresponds to a piston on a squared segment, when the 3 others are set to 0. Eq.\ref{eq:RMS} allows to identify normalization coefficients in Eq.\ref{eq:modalBasis}.

Here we allow ourselves a slight lack of mathematical rigour in that the integration bounds, on account of the removed portion shown in Fig.\,\ref{fig:pup} would not be exactly $1$ and $-1$. Since the amount of amplitude diversity needed to lift the even mode sign ambiguity is an open optimisation parameter, we gloss over this detail and assume the same normalisation for each and all segments. 

The WFE definition for segmented pupils differs from the definition over monolithic pupils as shown in Eq.\ref{eq:WFE}. 

\begin{equation}
\label{eq:WFE}
 WFE(\{c_{i}^{k}\}) = \sum_{k=1}^{4} \sqrt{ \sum_{i=1}^{N}{c_{i}^{k}}^{2} }
\end{equation}

\begin{figure}[h!]
\begin{center}
\captionsetup{font=small}
\subfloat[PSF diffraction limit]{\includegraphics[height=2.7cm]{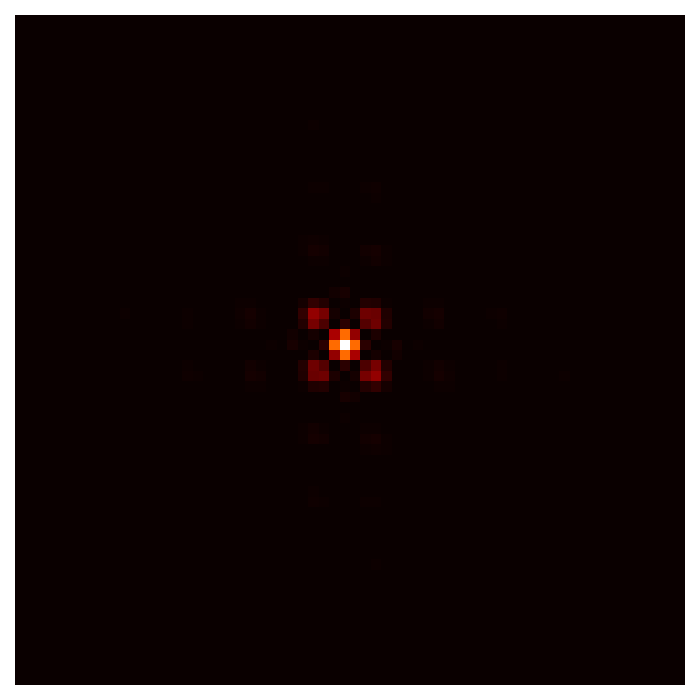}} 
\subfloat[PSF $WFE_{PTT}$=97nm]{\includegraphics[height=2.7cm]{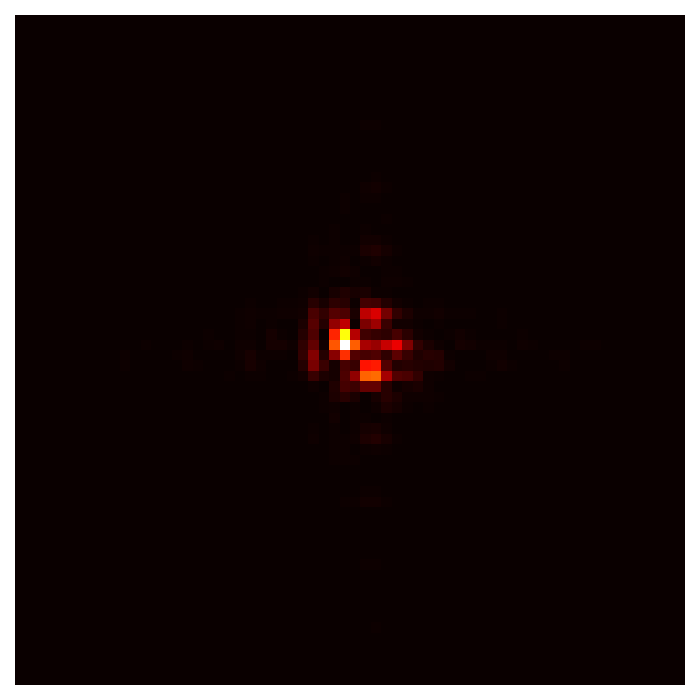}}
\subfloat[PSF $WFE_{PTT}$=454nm]{\includegraphics[height=2.7cm]{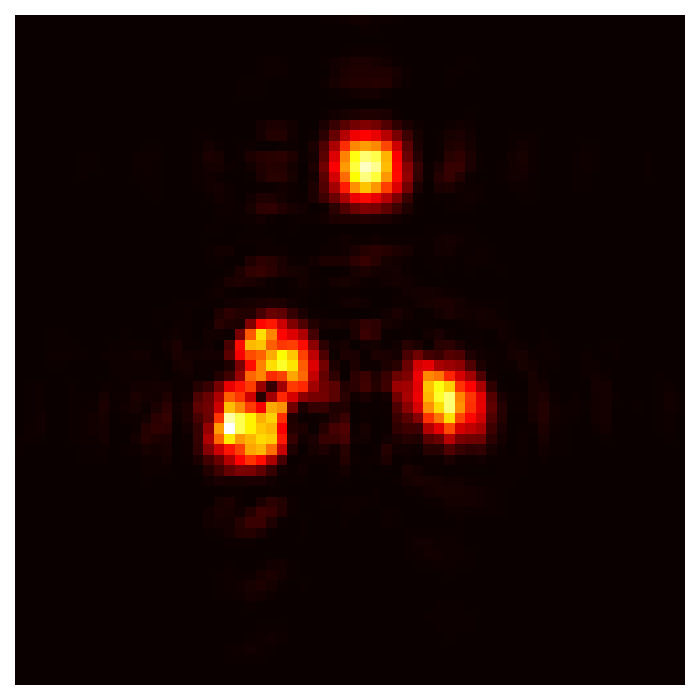}} \\
\subfloat[PSF diffraction limit + 30nm HO]{\includegraphics[height=2.7cm]{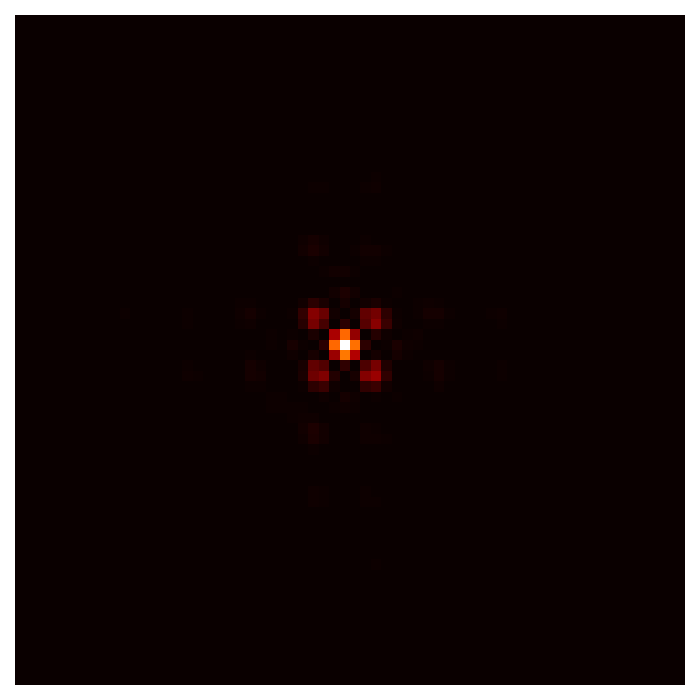}} 
\subfloat[PSF $WFE_{PTT}$ = 96nm $WFE_{HO}$=30nm]{\includegraphics[height=2.7cm]{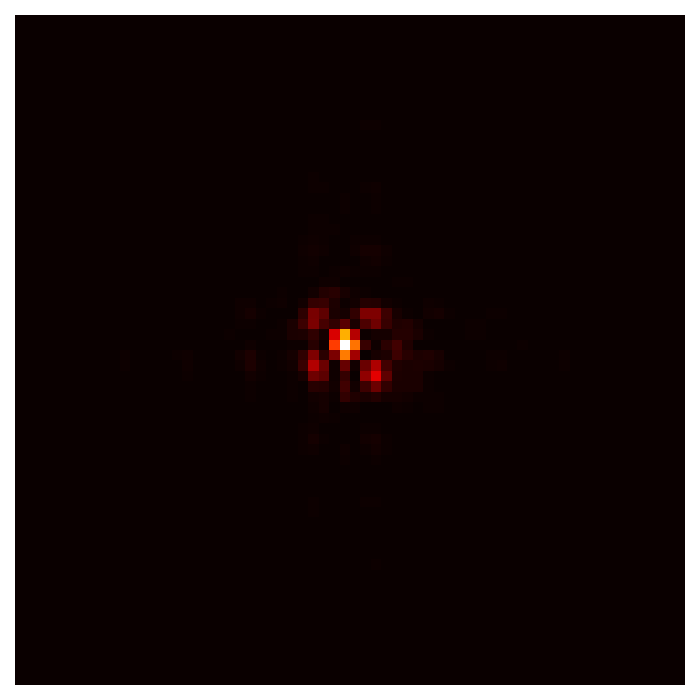}}
\subfloat[PSF $WFE_{PTT}$ = 97nm $WFE_{HO}$=60nm]{\includegraphics[height=2.7cm]{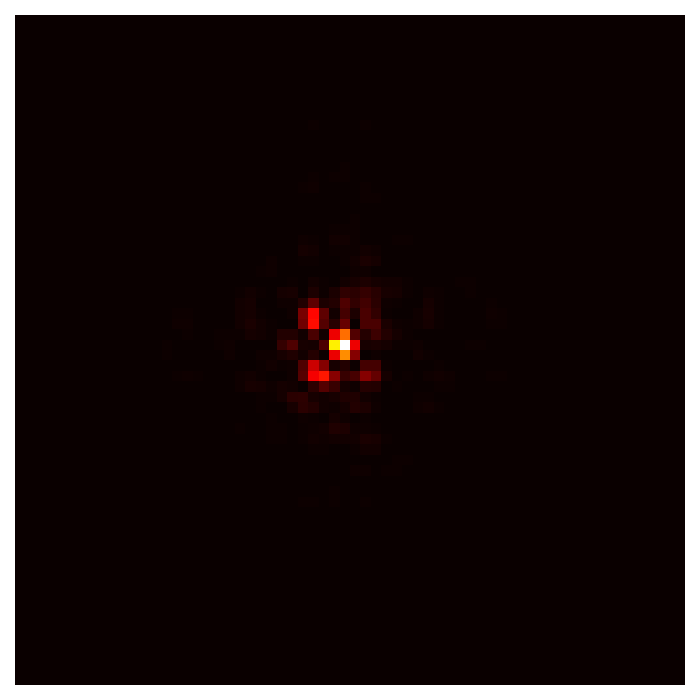}} \\
\subfloat[PSF diffraction limit, SNR = 10]{\includegraphics[height=2.7cm]{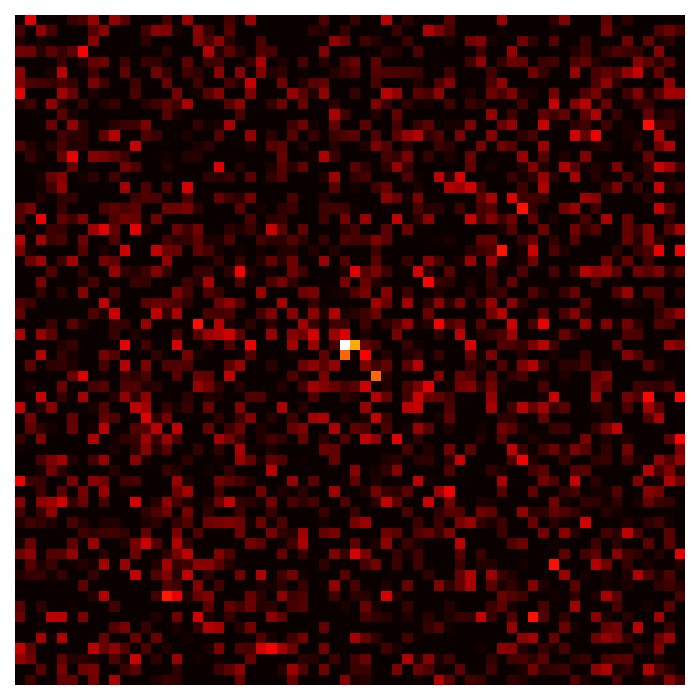}} 
\subfloat[PSF diffraction limit, SNR = 30]{\includegraphics[height=2.7cm]{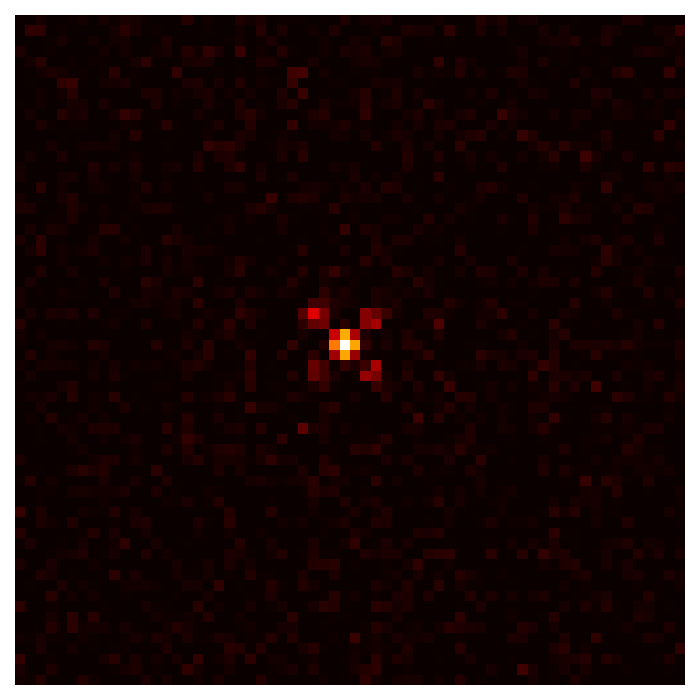}}
\subfloat[PSF diffraction limit, SNR = 100]{\includegraphics[height=2.7cm]{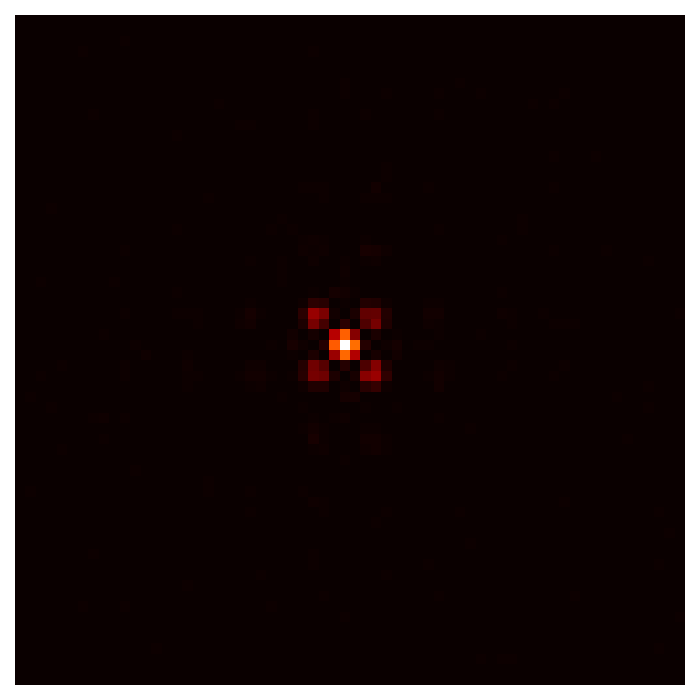}}
\caption{Illustration of AZIMOV PSFs obtained in different configurations of WFE and Signal to Noise Ratio (SNR). First row is PSF disturbed by PTT aberrations only. Second row is PSF disturbed both by PTT and HO aberrations. Last row is the noise influence depending on the SNR.}
\label{fig:PSFs}
\end{center}
\end{figure}

This WFE is described by a quadratic sum of the phase coefficients over each segment. Provided the segments are non-overlapping, the total RMS is the sum of individual RMS values and not their quadratic sum.\\
An example of the PSFs studied is shown Fig\,\ref{fig:PSFs}. These PSFs, paired with their Zernike coefficients, are directly used for the NN training.

\def\figsize{0.79}

\subsection{Data generation}\label{subsec:datagen}
\begin{figure}[h!]
\begin{center}
\includegraphics[height=4.5cm]{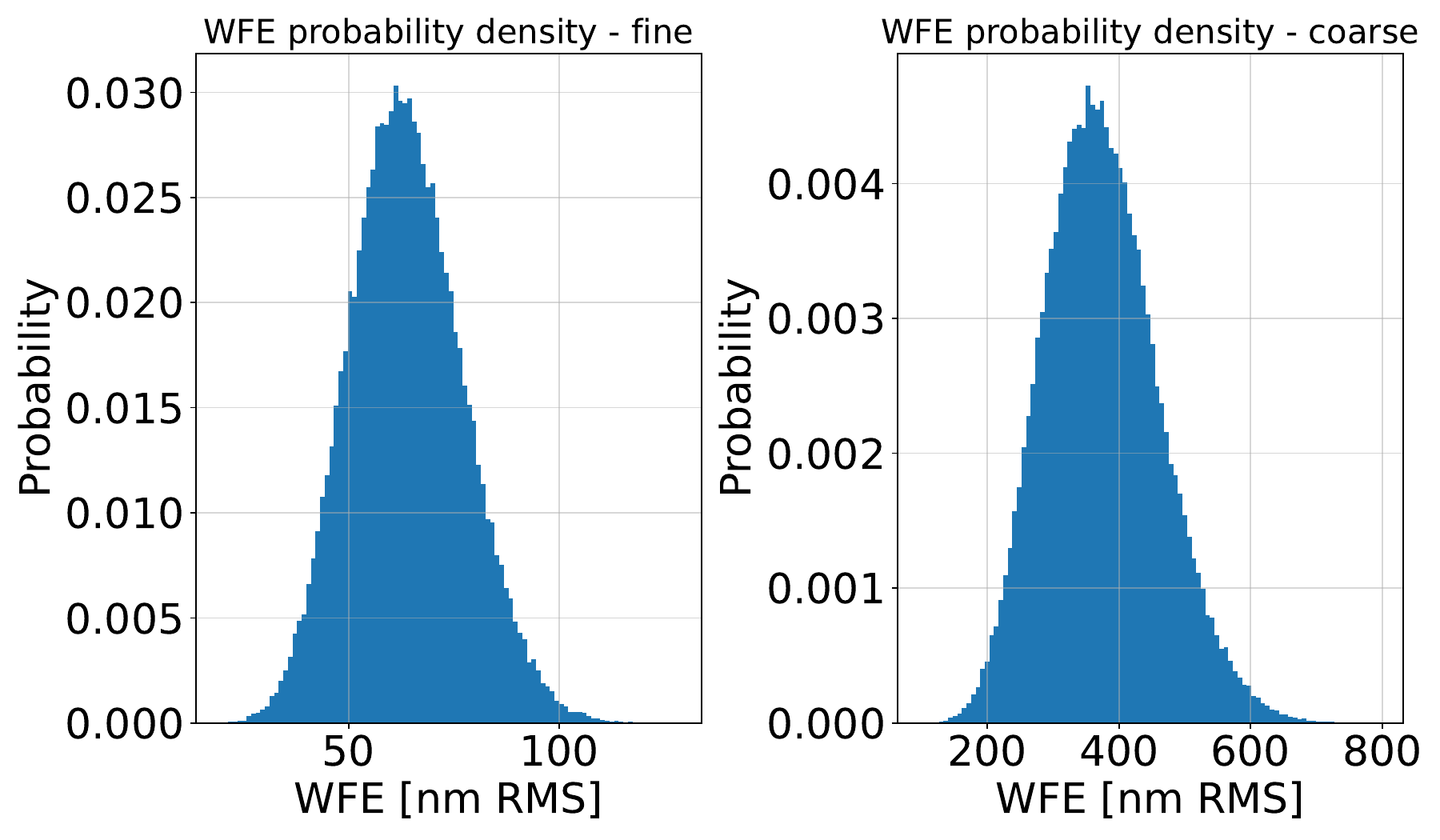}
\caption[example] 
{   \label{fig:WFE_Dist} 
Probability density of WFE over the full pupil for the training set calculated over $10^6$ samples using Eq.\ref{eq:WFE}. The figure on the left corresponds to the fine phasing. On the right, the distribution of the coarse phasing, studied in section \ref{sec:AZI}.}
\end{center}
\end{figure}

The training of NNs necessitates a substantial volume of data to effectively learn from diverse realizations. Unfortunately such an amount of real data isn't available yet, therefore data are simulated. All the generated imagery data are Nyquist sampled with a Field of View (FoV) of 64*64 pixels at $\lambda=800\unit{nm}$. We selected this Field of View (FoV) based on the anticipated amplitude of the Wavefront Error (WFE) to be measured.  The PSFs  global intensity is normalised to 1. Fig.\,\ref{fig:WFE_Dist} presents the two studied distributions, one for large WFE errors corresponding to the initial mirror segments at the deployment  (i.e., coarse phasing), and the other for smaller WFE error for the final fine phasing step. The distribution of the 12 piston-tip-tilt coefficients follows a Gaussian distribution centered on $0\unit{nm}$ with $\sigma_{fine} = 10\unit{nm}$ or $\sigma_{coarse} = 70\unit{nm}$. However, identification of piston, under a single-image FPWFS setting, is constrained to a range between $-\frac{\lambda}{4}$ and $\frac{\lambda}{4}$ or their estimate can suffer from the so-called lambda ambiguity. We constrained piston values on a $\frac{\lambda}{2}$ width interval so that no piston can be identified at +$\frac{\lambda}{2}$ while another would be identified as -$\frac{\lambda}{2}$ since it will lead to the same PSF. Therefore, for the coarse phasing, the piston values are randomly drawn following the same distribution than for the fine phasing ($\sigma_{coarse}^{piston} = \sigma_{fine}^{piston}$) so that the same NN model studied in the following section can be used also at Part.\,\ref{sec:AZI}.The capture range of the depicted distribution enables a range of minimum and maximum values within approximately [-50, 50] nm RMS, corresponding to [-200, 200]\,nm\,Peak to Valley (PV) for the piston and [-350, 350]\,nm\,PV for tip-tilt. This amplitude is reached after an initial phasing step using ELASTIC method$\,$\cite{vievard2017large}, or our embedded onboard capacitive displacement sensors$\,$\cite{Schwartz}.

The determination of an optimal dataset size, depending on the complexity of the problem and the complexity of the learning algorithm (e.g., a nonlinear relationships between input and output features), is controversial as there is currently no definitive methodology for its precise determination. We generated a dataset comprising $10^5$ samples randomly split: 90\% of the samples are allocated to the training set, while the remaining 10\% are assigned to the validation. An additional $10^4$ samples are generated exclusively for the testset so that the NN model can be test with unseen data. All the datasets follow the same coefficient distribution.

\subsection{Disturbance sources}
    \subsubsection{Noise}
In a low flux regime, the images mostly suffers from readout noise, estimated at 2.1 electrons (e.g. using a Sony Pregius-S IMX530 CMOS detector with 5320x4600 pixels) whereas photon noise dominates on high flux cases. These are the two noises considered in the study, while other sources of noise (background, gain, drift) have been chosen to be ignored. Therefore, we define the Signal to Noise Ratio (SNR) such as

\begin{equation}
\label{eq:SNR}
    SNR = \frac{f}{\sqrt{f + \sigma_{ron}^2}}   
\end{equation}
where $f$ is the star flux in numbers of photons and $\sigma_{ron} = 2.1 \unit{e^-}$ the readout noise.  The photon noise is described by a Poisson distribution $\mathcal{P}(f)$ for each pixel. The readout noise is a Gaussian noise drawn at $\mu = 0, \sigma = \sigma_{ron}$ added to the pixel value.
The corresponding noisy images can be found Fig\ref{fig:PSFs}.

\subsubsection{Higher-order aberrations}
    Higher-order aberrations introduce additional distortions to the wavefront on top of the PTT. However their impact is different from the PTT: although they degrade the PSF and therefore the delivered resolution, they can not be compensated for. We carry out an analysis of their impact in the system: how they degrade the PTT estimation, and if it is needed to include them in the training set for PTT. These aberrations are added in addition of the PTT in the wavefront. They follow a $1/f^2$ power spectrum, typically encountered with good quality optics$\,$\cite{dohlen2011sphere} and are orthogonal to PTT aberrations, see part\,\ref{sec:IF}. The magnitude of HO aberrations is approximately $30\unit{nm\ RMS}$ over the full pupil, or in other words $\sum{1/f^2} = 30\unit{nm\ RMS}$, consistent with current estimates for AZIMOV.

\section{Methods}\label{sec:Methods}
    \subsection{Deep Learning}
  Deep Learning is a type of machine learning method able to extract representations or model directly from data with a complex architecture of learning successive layers with a high level of abstraction (increasingly meaningful representations). These architectures leverage a combination of diverse non-linear transformations to accomplish their objectives. The NN architecture has a lot of influence on the output accuracy and is composed of different layers, of which the 2 primary types are:

\begin{itemize}
    \item The \textit{fully connected layers}$\,$\cite{goodfellow2016deep}. A layer composed of neurons. A neuron computes a weighted sum of its input, plus a bias and then applies a non-linear activation function. The model parameters are therefore the neurons weights and biases.
    
    \item The \textit{convolutional layer}$\,$\cite{goodfellow2016deep}, particularly adapted for image processing, computes the convolution of an image with one or several kernels. In this case, the neural network optimizes all the convolution kernels and biases.
\end{itemize}

\begin{figure}[h!]
    \begin{center}
        \includegraphics[height=2.9cm]{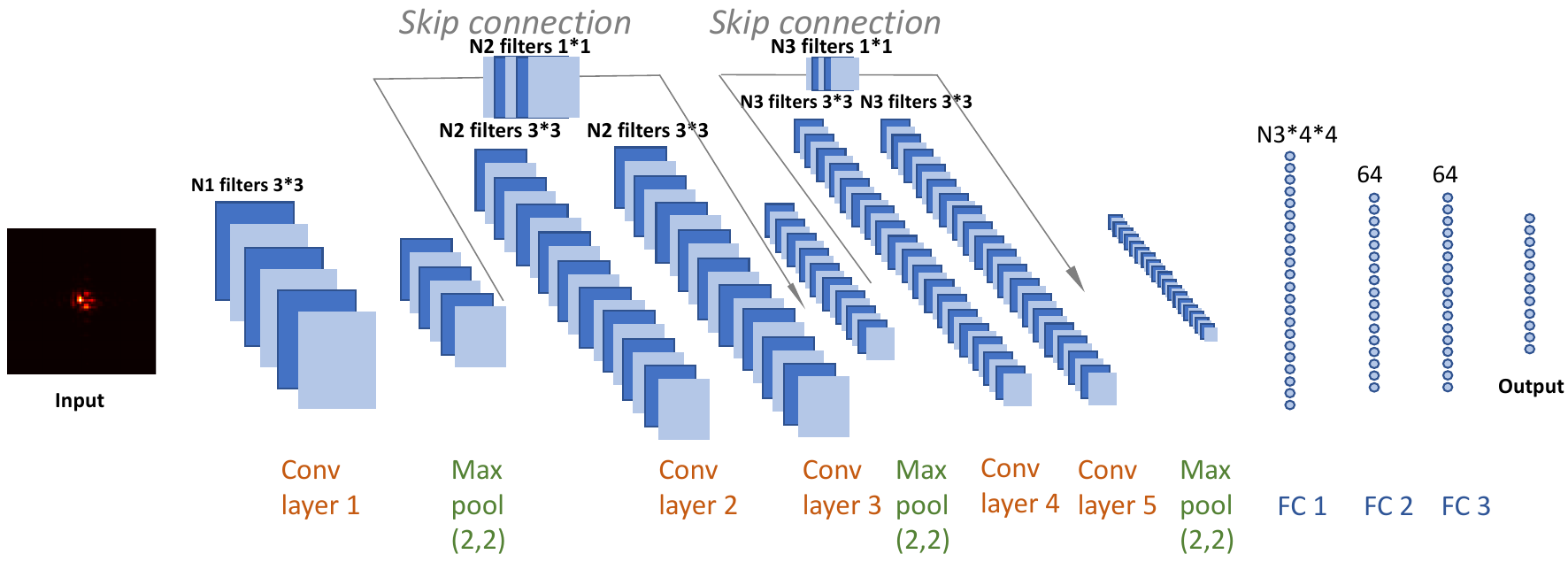}
        \caption[example] 
        {   \label{fig:archi} 
        Adopted Resnet architecture.}
    \end{center}
\end{figure}

The model architecture is created by adding layers one to another. Choosing the right architecture depends on the more important criterion of the project: it can be either the accuracy of the estimation, the model size or the inference speed. In our work we aim to find the right balance between performance (the precision of WFE estimation) and model size (the size of the NN algorithm which needs to be embedded onboard the CubeSat). The number of model parameters depends on the choice of NN architecture. Decreasing it avoids overfitting and makes the model size smaller. Since our input data consists of 2D images, convolutional NNs have fewer parameters compared to fully connected layers and can capture pixel neighborhood information effectively. The simplest NN architecture to infer coefficients from an image is the VGG-Net$\,$\cite{simonyan2014very} or the ResNet$\,$\cite{he2016deep} architecture that are optimal in terms of number of parameters. Historically, the Resnet architecture is known for the "\textit{skip connection}", allowing the loss gradient to float easily through the deepest layer of the model during backpropagation.
To achieve the best trade-off between performance and time complexity, we consider these two highly suitable architectures: The VGG-net$\,$\cite{simonyan2014very} and the Resnet$\,$\cite{he2016deep}. \\
We therefore propose a NN baseline composed of 5 convolutional layers stack as VGGnet architecture  (2 convolutional layers followed by 1 max-pooling layer) and 3 fully connected layers with ReLu activation function. In the case of the Resnet, two Skip Connections are added to the architecture (each of the skip connection is a convolutional layer with a 1x1 filter) : the first one starts after the first layer to merge with the output of the 3rd layer, and the second one after the 3rd layer to merge after the 5th layer. The architecture is presented Fig.\,\ref{fig:archi} and has been chosen in order to find the smallest architecture suitable for reaching the diffraction limit requirement. Indeed such architecture goes up to $10^6$ number of parameters whereas the smallest Resnet found in the literature, the Resnet-18, reaches $10^7$ parameters, 10 times more than our heavier network.

N1, N2 and N3 corresponds to the number of filters of the convolutionnal layers. This number of filters impacts the internal parameters and changes the size of the NN model (see Tab\ref{Tab}), all model parameters are stored in a double precision float64 format as it is the the default format in pytorch library. The number of parameters also impacts the computation time and influences the model's performance. Increasing the number of parameters can improve performance until the point of overfitting occurs, leading to limited generalization on new data.

\begin{table}
\addtolength{\tabcolsep}{-1pt}

\begin{tabular} {||c | c c c c c||}
 \hline
 Label & N1 & N2 & N3 & \# parameters & Model size [MB] \\ 
 \hline\hline
 NN1 & 4 & 8 & 16 & $13.10^3$ & 0.134 \\ 
 \hline
 NN2 & 8 & 16 & 32 & $31.10^3$ & 0.278 \\
 \hline
 NN3 & 16 & 32 & 64 & $94.10^3$ & 0.94 \\
 \hline
 NN4 & 32 & 64 & 128 & $326.10^3$ & 2.6 \\
 \hline
 NN5 & 64 & 128 & 256 & $1.2.10^6$ & 11 \\ [1ex] 
 \hline

\end{tabular}
\caption{Resnet model sizes depending on the number of convolutional filters. This table approximates  VGG-net sizes as well, since Resnet architecture has only slightly more internal parameters due to the skip connection that needs to link the number of filters from the input layer to its output layer.}
\label{Tab}
\end{table}

The training step relies on a minimization algorithm of a cost function commonly known as Loss function. Conceptually, a loss function is a way of prioritizing which error to fix from our training samples, so that NN parameters updates result in adjustment to the output, decreasing the loss. The minimization algorithm optimizes the NN parameters (weights, biases and filters) to minimize the Loss Function.  Usually, the Root Mean Square Error is used  but in our study, the WFE over the pupil is computed as the loss function:

\begin{equation}
\label{eq:WFE_error}
    Loss_{WFE} = WFE(c_{i}^{k}-\hat{c}_{i}^{k})
\end{equation}
where $c_{i}^{k}$ is the true coefficient for mode $i\in [1..3]$ and segment k and $\hat{c}_{i}^{k}$ is the estimated coefficient. This loss, as mentioned above, allows the model to adjust accurately the model parameters according to the optical criterion on the pupil. \\

\subsection{Classical phase-diversity and image sharpening}\label{subsec:class_method}

Other more classical FPWFS methods have also been propose for the phasing of segmented telescopes, namely phase diversity and image sharpening. The whole problem of phase diversity (PD) lies in the estimation from the focal plane image and the defocused image of the unknowns that are the phase and the object$\,$\cite{blanc2002identification}. However, in our case, the object isn't to be estimated as it's a point source. Since our system does not include any additional optical components and aims to maintain the integrity of the focal plane image without any defocus, it is not possible to acquire a defocused image. Therefore, we propose a single-image phase diversity that minimizes the distance between the real PSF and the fitted one by adjusting the 12 phase coefficients. A criterion based on a quadratic pixel-to-pixel difference between the real PSF and the fitted one is minimized iteratively using Conjugate-Gradient method$\,$\cite{polyak1969conjugate}. A Tikhonov regularization is added to the criterion, in order to avoid the divergence of the phase coefficient, particularly at low SNR.  Conjugate-Gradient has shown the best performance in time, number of iteration and WFE among others minimization methods.

Another method is Image Sharpening (IS). By taking focal-plane images, the algorithm iteratively maximizes an image maximum intensity centered window  by adjusting iteratively the PTT. Classical metrics are the maximum intensity in the center of the image, contrast, the sum of the pixel squared or alternative optimisation method are also possible and may lead to faster and improved estimates. Powell optimization methods$\,$\cite{powell1994direct} has shown the best efficiency in terms of noise robustness and computing time. To perform IS, we consider an exposure time of $0.1\unit{ms}$. This exposure time is taking into account in the method calculus time as an image has to be acquired at each iteration of the algorithm. The computation time of generating the PSFs is deduced from the total computing time.

\section{Sample numerical results} 
\label{sec:Res}

    \subsection{Model architecture}\label{subsec:FullPup}

First,  both types of model architectures (VGG-net and Resnet) are trained using identical hyperparameters (\textit{batch size}, \textit{learning rate}, \textit{epochs}) and their inference performances are compared. The performances of the Resnet architecture are observed to systematically outperform the VGG-net even after fine-tuning both model hyperparameters (i.e. other models are trained) enhancing performance with respect to the architecture and number of parameters. The optimal set of hyperparameters for the Resnet includes a progressive scheduled Learning Rate (LR) depending on the epoch \textit{e}: 

$$
LR(e) = \left\{
    \begin{array}{ll}
        10^{-4} & \mbox{if}\ e<70 \\
        10^{-6} + (10^{-4} - 10^{-6})exp^{-(e-70)*0.0125} & \mbox{otherwise.}
    \end{array}
\right.
$$

Specifically, a batch size of 32 PSFs is used during 300 learning epochs.

\begin{figure}[h!]
\begin{center}
\includegraphics[height=4.5cm]{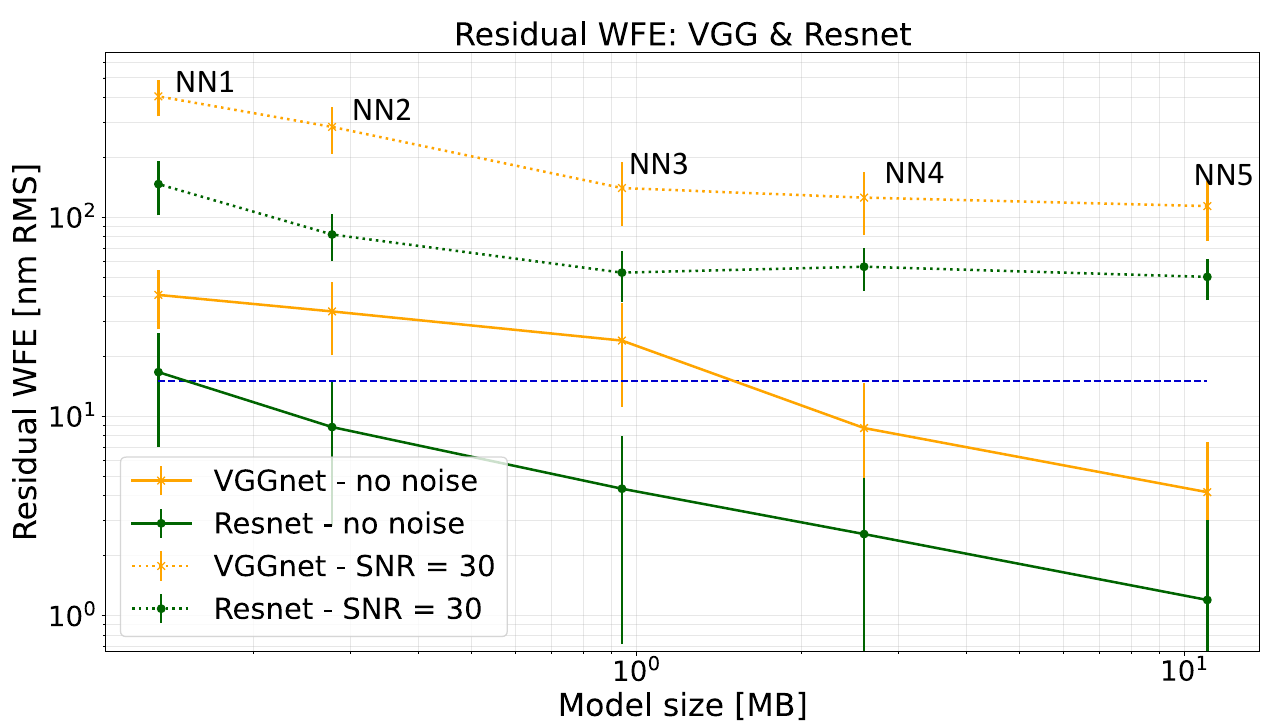}
\caption[example] 
{   \label{fig:vgg_resnet} 
Performance comparison of estimation between the VGG-net architecture (yellow line) and the Resnet architecture (green line) as a function of the model size. Dotted-line corresponds to models estimation for data at SNR = 30. The dashed blue line represents the residual target to reach the diffraction limit at $15\unit{nm\ RMS}$. VGG-net size are rounded at the same value as Resnet for the graphics readability.
}
\end{center}
\end{figure} 

Figure \ref{fig:vgg_resnet} presents the performance comparison between the two architectures mentioned above for various model sizes. Performance is compared using the same test dataset and optimal sets of hyper-parameters, in terms of \textit{residual WFE}, assuming a perfect correction based on the algorithm estimates. Both NN models are trained with noiseless data and tested with either noiseless data or data at SNR = 30 (the noise study will be developed in part\,\ref{subsec:noise}).
As expected, larger models lead to better performance, but at the cost of increased computational complexity. The best performance reaches about $1\unit{nm}$ of WFE, far below the residual target at $15\unit{nm}$. However, when facing a low SNR, both NN models show weak performance where the best performance achieved is $50\unit{nm}$ of WFE for the heaviest Resnet. Improvement on noisy data are shown in part\,\ref{subsec:noise}.
Whatever the model size, the Resnet outperforms the VGGnet. Additionally, Resnet has demonstrated faster convergence and lower loss function values compared to VGGnet. The Skip Connection in the Resnet architecture enables more effective parameter updates in the top layers and accelerate the convergence of the loss function. Based on the NN performance, the Resnet architecture is preferable. The requirement of $15\unit{nm\ RMS}$ is fulfilled by nearly all Resnet models regardless of the model size. However, in presence of noisy data, the methodology should be improved to reach the diffraction limit.\\ 
Subsequently, only the Resnet architecture will be examined in the rest of the paper. It is important to note that the simulated data used for evaluation lacks noise diversity and higher-order aberrations. Therefore, it is crucial to assess the robustness of the models by introducing noise and HO aberrations into the samples.

    \subsection{Models robustness to noisy data}\label{subsec:noise}
As mentioned above, images suffer from detector and photon noise. For the training step, the noise is added "on-the-fly" meaning that during each epoch, the noise is added to each noiseless PSF so that the noise realisation is always different at each epoch for each PSF. Two scenarios are identified: 
\begin{itemize}
    \item Scenario 1: Each NN model is trained with a noiseless dataset and is tested with a noisy dataset. Each single NN model is then tested over different SNRs. Several model sizes are tested. The aim is to evaluate the error made on noisy datasets by a model that has been trained with a noiseless dataset. 
    \item  Scenario 2: Each NN model is trained with noisy samples at a given SNR. Each NN model is then tested on noisy data at the same SNR than for the training. 2 NN model sizes are tested. The aim is also to evaluate the error made on noisy datasets when the NN models have been trained with noisy datasets.
\end{itemize}

\begin{figure} [h!]
    \begin{center}
        \includegraphics[height=4.5cm]{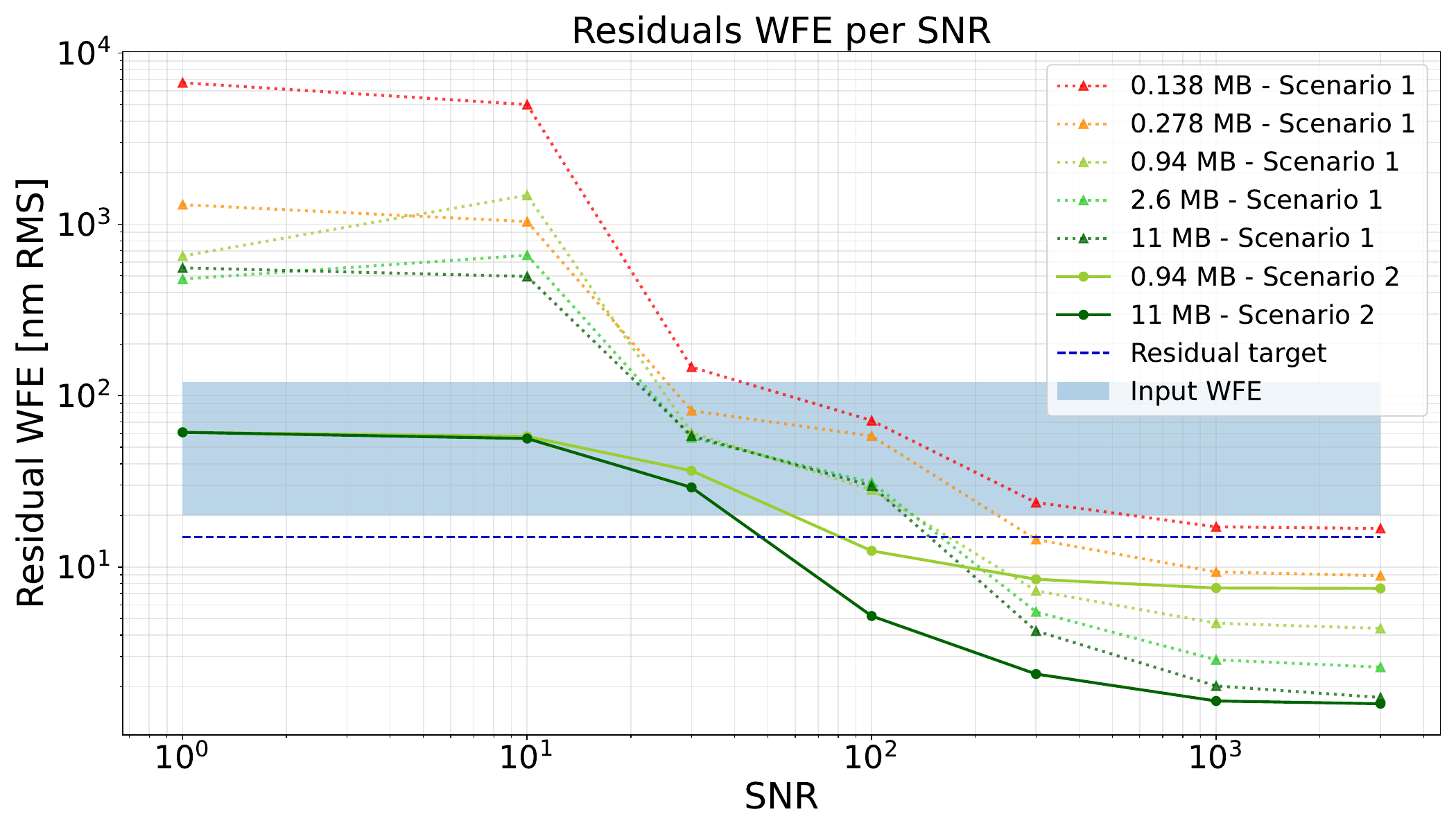}
        \caption[example] 
        {   \label{fig:noise} 
            Resnet robustness to noise: scenario 1 (dotted lines) and scenario 2 (solid lines). For the scenario 2 curves, each point represents one NN model, trained and tested on the corresponding SNR.
        }
    \end{center}
\end{figure} 

We conducted a performance comparison between the two scenarios based on residual WFE after correction. Fig.\,\ref{fig:noise} illustrates the residual wavefront error depending on the SNR. The dotted lines represent NN models trained without noise, while the solid lines represent NN models trained with noisy PSFs (each point on the solid line corresponds to an NN model trained for a specific SNR). The performance of the models unsurprisingly improves as their size increases. The diffraction limit, defined by WFE $<$ \textit{Residual target}, is not achieved by the smallest models (dashed red line) even at high SNR.
For low SNR, where the detector noise prevails over the photon noise, the \textit{scenario 1} outputs an estimation error larger than the initial WFE: this would lead to an increase of the WFE if used in closed loop. On the other hand, the \textit{scenario 2} networks estimation stays in the mean of the input WFE distribution: during training, the models minimize their loss function by setting all the phase coefficients to 0 as the PSF geometry cannot be properly identified in the noise.
At high SNR, in both scenarios, the largest model (NN5) converge towards the same WFE value. However, the NN3 \textit{scenario 2} model exhibits a slight limitation, achieving a WFE slightly below $10\unit{nm\ RMS}$, whereas the same model trained without noise performs better. \\
An alternative approach is to train models using a range of SNRs drawn uniformly in the interval [30, 100] to assess their performance.  

\begin{figure} [h!]
    \begin{center}
        \includegraphics[height=4.5cm]{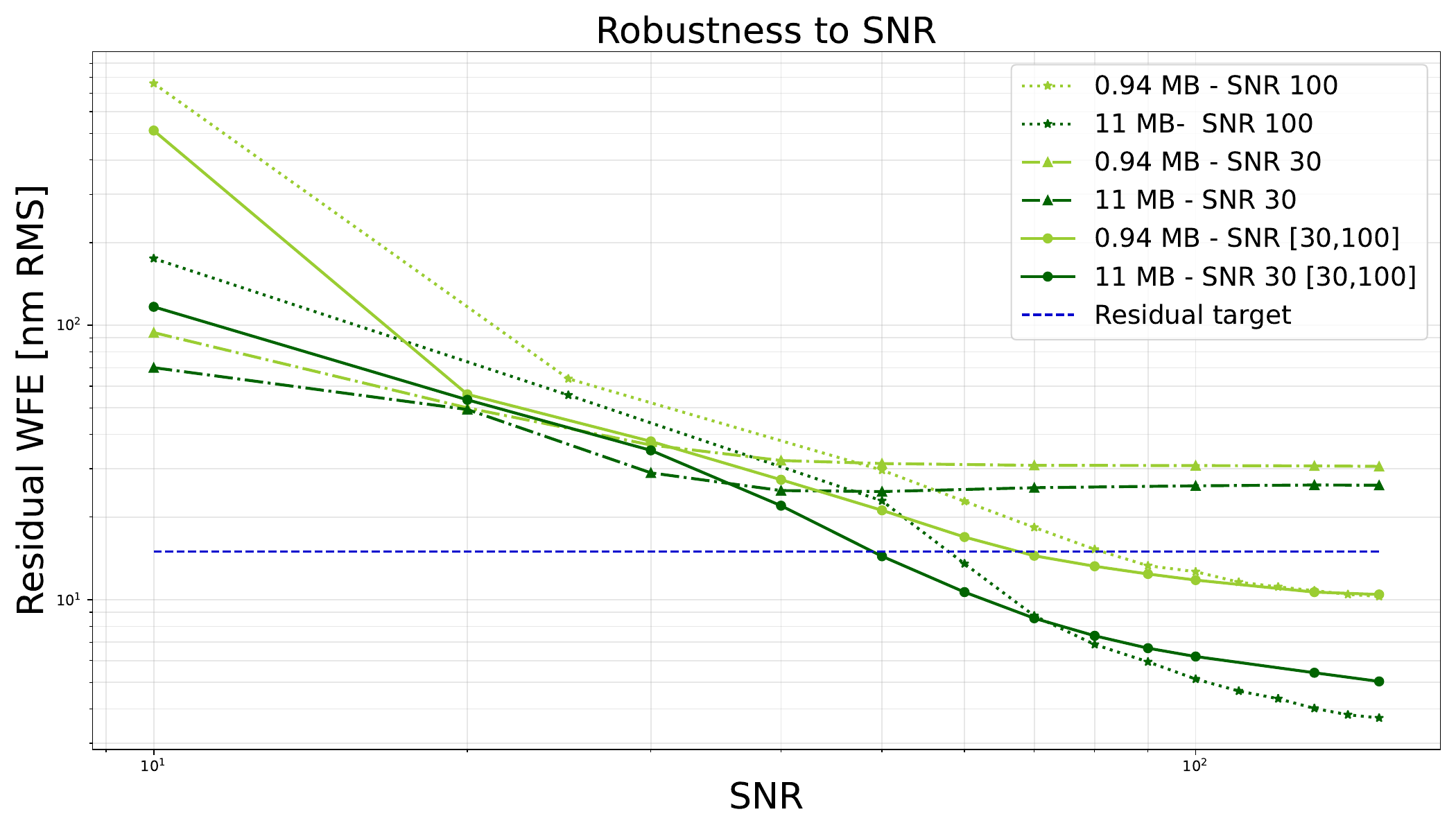}
        \caption[example] 
        {   \label{fig:noise_rob} 
            Resnet robustness to SNR. Dark green curve illustrates the estimation quality of the NN5, light green shows NN3 performances.
        }
    \end{center}
\end{figure} 

Fig.\,\ref{fig:noise_rob} illustrates a comparison of wavefront estimation quality between models trained on a fixed SNR and those trained on varying SNRs. Remarkably, the models trained on various SNRs show great performance, reaching the diffraction limit more rapidly at SNR=50 for the NN5 and SNR=70 for the NN3. Surprisingly, even at high SNRs, the models trained on varying SNRs demonstrate comparable performance to the models trained at fixed SNR. However, in a very low flux regime, performance are worsening the wavefront. Globally speaking, an SNR of at least 50 is required to reach the diffraction limit with a relatively small NN model. This is somehow similar to Phase diversity. \\

    \subsection{Models robustness to higher order aberrations}
 To validate the performance of the neural network, the model is also trained considering the presence of higher-order aberrations on the primary mirror. In the same way than for the noise robustness discussed in subsection \ref{subsec:noise}, we distinguish between two scenarios:

\begin{itemize}
    \item Scenario 3: Each NN model is trained with a dataset where higher-order aberrations are not considered, and is tested on PSFs generated with PTT and higher-order aberrations.
    \item  Scenario 4: Each NN model is trained with samples generated from PTT and higher-order aberration phase map. The model is then tested over data that suffer from the same higher-order amplitude.
\end{itemize}
For these two scenarios the 12 PTT coefficients follow the same distribution. They are the only coefficients inferred by the NN model since the HO aberrations cannot be compensated for. The \textit{Residual WFE} refers here to the PPT part of the phase (i.e. excluding HO aberrations).

\begin{figure} [h!]
    \begin{center}
        \includegraphics[height=4.5cm]{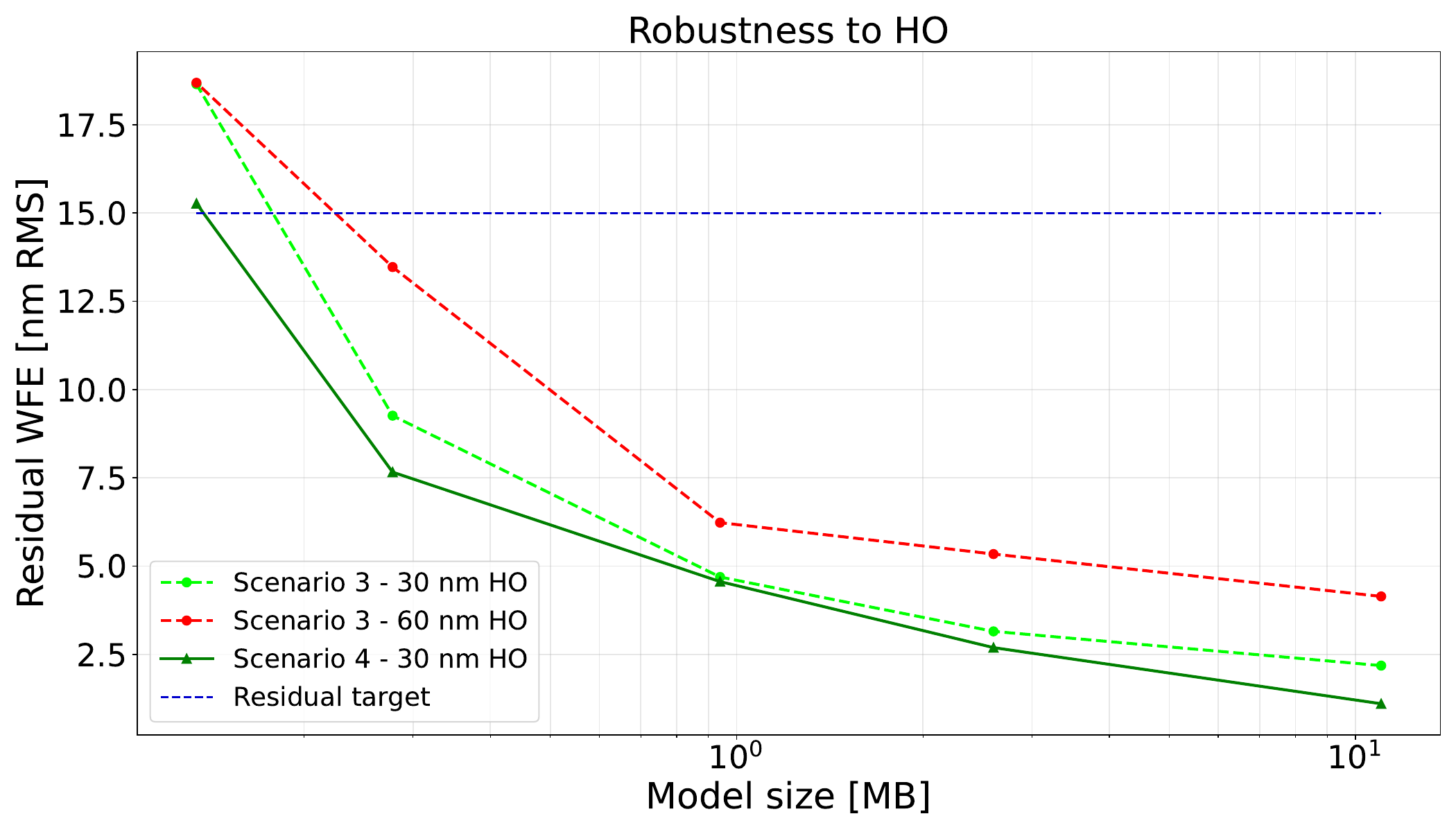}
        \caption[example] 
        {   \label{fig:ho} 
                Resnet performance as a function of model size for images including higher-order aberrations. The dashed light green line corresponds the scenario 3 (i.e., no HO aberrations included in the training) for $30\unit{nm\ RMS}$ of HO. The red line corresponds to the scenario 3 for $60\unit{nm\ RMS}$ of HO aberrations. In solid green line, scenario 4 results for a NN trained with PSF aberrated with PTT and HO.
                }
    \end{center}
\end{figure}

Fig.\,\ref{fig:ho} shows that even in the presence of unexpected HO aberrations, the inference step achieves a performance better than the diffraction limit requirement. The difference between \textit{Scenario 3} and \textit{Scenario 4} is only of a few nanometers, indicating that all model size exhibits robustness against small changes in the focal plane image due to HO aberrations. Surprisingly, the most robust model is for a size of $1\unit{MB}$, where the difference between the \textit{Scenario 3} and the \textit{Scenario 4} is the smallest, approximately $0.3 \unit{nm\ RMS}$. In contrast the heaviest model begins to overfit the data: it infers with the best accuracy the raw data but when new data is introduced for testing, the performance slightly degrades by a few nanometers, while still reaching the diffraction limit. The inference accuracy worsens when stronger HO aberrations are added to the phase map: as shown by the red curve Fig.\,\ref{fig:ho} which represents the inference performance when $60\unit{nm\ RMS}$ of WFE are added on top of PTT. However, the performance continues fulfilling the accuracy requirement and adding $60\unit{nm\ RMS}$ on the phase map introduces only a few nanometers of estimation errors.

\subsection{Comparison to classical minimization methods}
Having explored the robustness of different model sizes in term of estimation accuracy, we now turn our attention to comparing our method with Phase Diversity and Image Sharpening (IS) in terms of computation time and performance.
Phase diversity relies on an numerical iterative minimization of a quadratic criterion over the image pixels, involving a computing time depending on the number of iteration of the algorithm. Each iteration requires several Fourier Transform as the model PSF is fitted to the measured data (see Part.\,\ref{subsec:class_method}). Three tolerance criteria are chosen: one fast but less accurate $\epsilon^{PD}_1 = 10^{-4}$ one finer but slower $\epsilon^{PD}_2 = 10^{-7}$ and a very slow and precise $\epsilon^{PD}_3 = 10^{-15}$.  The principle is also iterative for IS: the metric is maximized until a tolerance criterion is reached, here a tolerance criterion of $\epsilon^{IS} = 2.10^{-4}$ is chosen. In our case, the metric employed is the maximum pixel intensity of a 2x2 window centered on the image. At each step, PTT corrections are applied (i.e. mirrors are moved), another image is acquired, and the maximum intensity is subsequently calculated.
 \\
Data used for the comparison are from the same dataset so that the same data are compared between the 3 methods.

\begin{figure} [h!]
    \begin{center}
        \includegraphics[height=4.5cm]{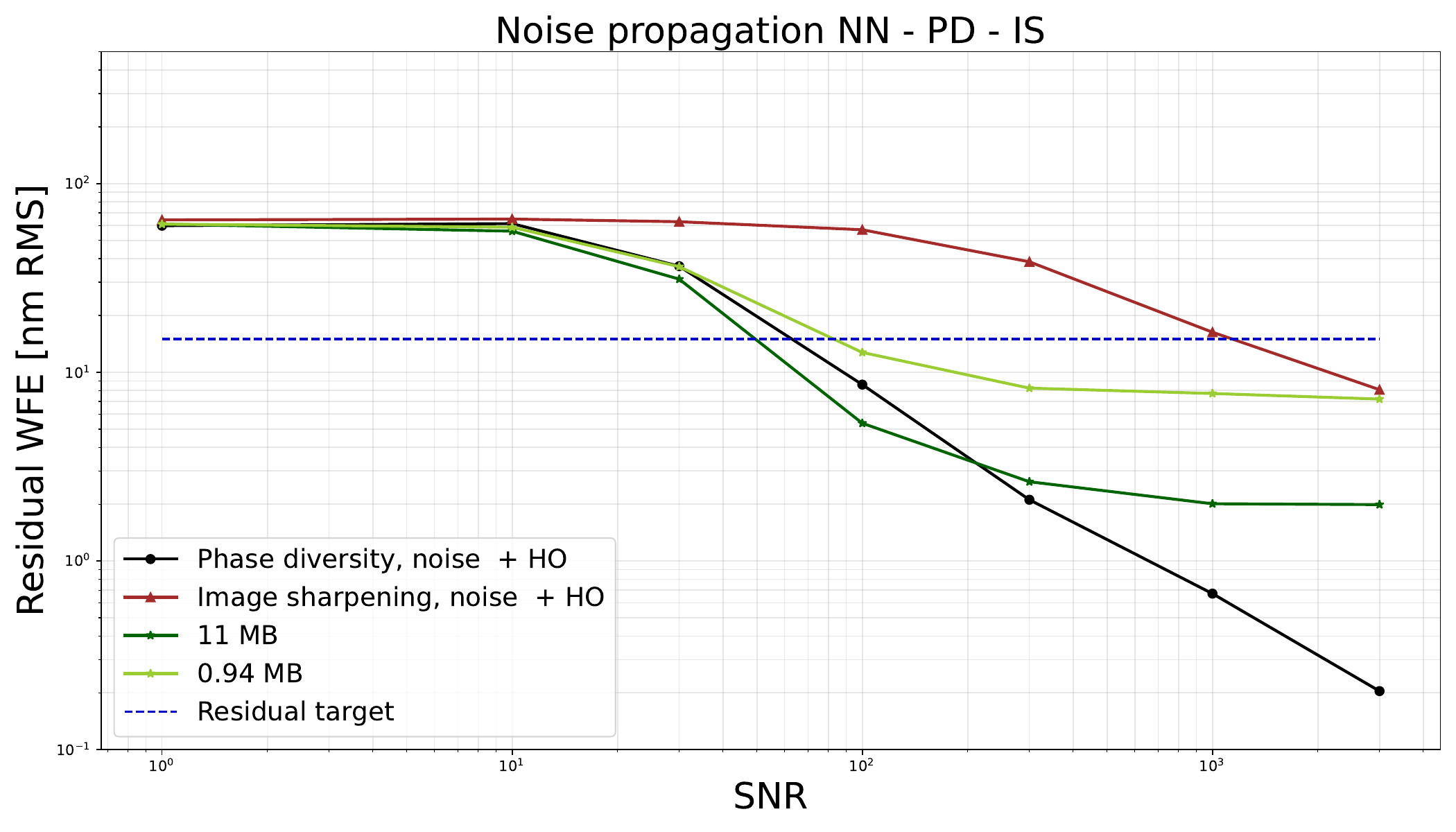}
        \caption[example] 
        {   \label{fig:ho_n_noise} 
           PTT estimation of the NN (green line), phase diversity method at $\epsilon^{PD}_3 = 10^{-15}$ (dark line) and Image sharpening (brown line) facing the same dataset of PSF generated with $30 \unit{nm\ RMS}$ of higher-order aberrations at several SNRs regardless of the computing time.
        }
    \end{center}
\end{figure} 

\begin{figure} [h!]
        \includegraphics[height=4.5cm]{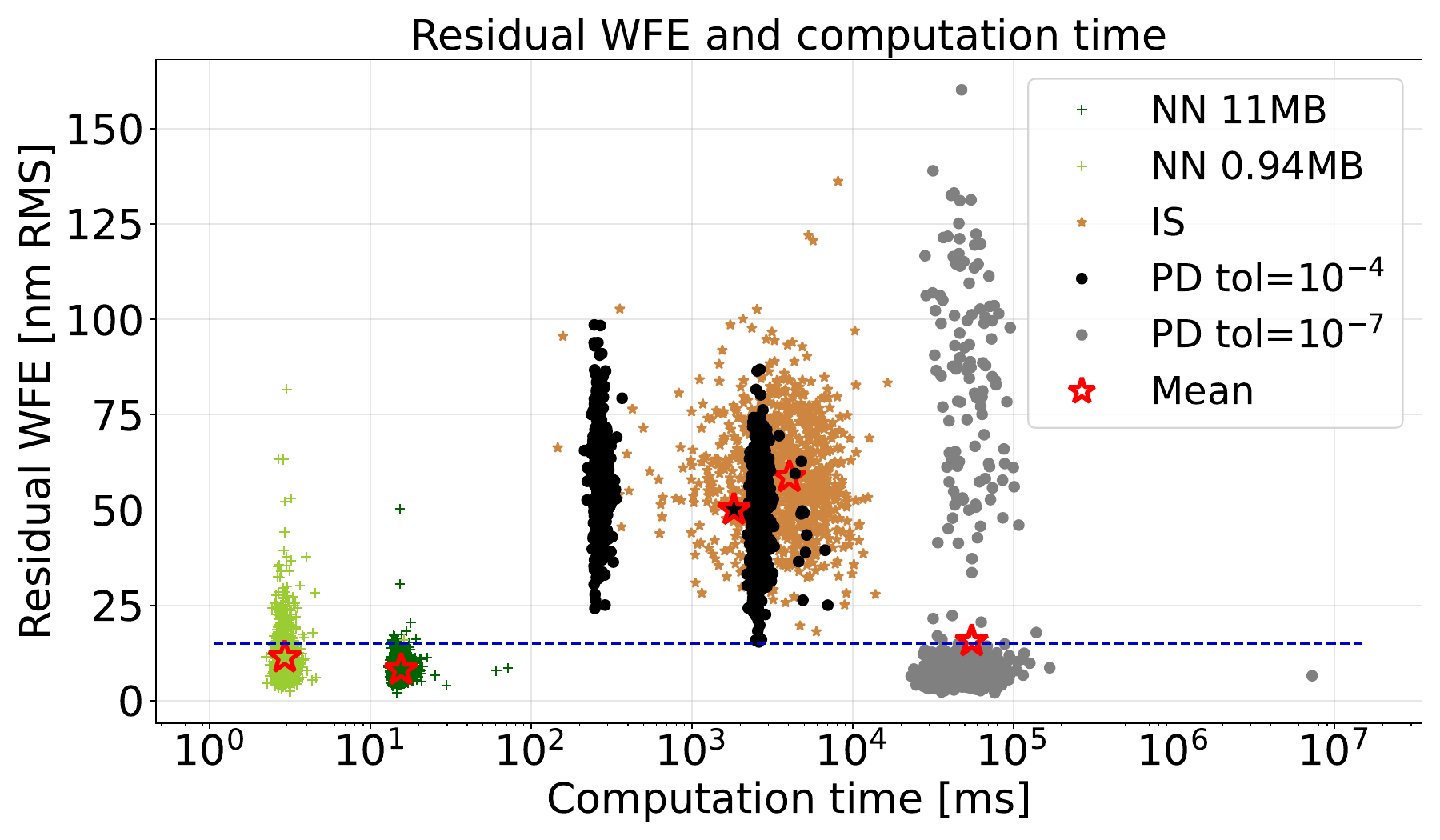}
        \caption[example] 
        {   \label{fig:time_wfe} 
           Estimation time and precision of estimation for NN, Phase Diversity (PD) and Image Sharpening (IS). Two types of PD are studied, one fast but less accurate, one slow and more accurate. Test made over the same 1000 samples from the NN testset with HO aberrations for an SNR=100. Algorithm computed on a 2,3 GHz Quad-Core Intel Core i7. 
        }
\end{figure} 

We aim to compare the three methods in terms of wavefront estimation performance under the presence of noise and higher-order aberrations. NN3 and NN5 are the two neural networks used to compare our 3 methods, trained on at the corresponding SNR with HO aberrated data.
NN coefficients estimation yields better phase estimations for both the NN5 and NN3 across all SNRs levels. Plus these models demonstrate a significant improvement in terms of computational time compared to PD an IS. NN models with their relatively lightweight architecture are much faster, with the majority of computational time occurring during training (which is obviously not taking into account for the inference).
Fig.\,\ref{fig:ho_n_noise} and \ref{fig:time_wfe} illustrate the performances of both methods relative to their execution time. The $\epsilon^{PD}_3 = 10^{-15}$  criterion and $\epsilon^{PD}_2 = 10^{-7}$  achieve nearly the same performance at SNR\,=\,100 but with a different computing time.  Fig.\ref{fig:time_wfe} only $\epsilon^{PD}_2$ is used to provide a fairer comparison between the methods. Compared to the NN inference, which is stable in terms of computing time, PD shows limitations in execution time. Firstly, PD takes longer than NN inferences, and secondly the standard deviation of the computing time is larger. In addition, 2 distinct clusters are identified computing phase diversity with a tolerance criterion of $10^{-4}$, indicating that the number of iterations to reach convergence depends on the criterion shape. Better estimations can be achieved with a convergence  criterion of $10^{-7}$. However, it requires a longer computing time. Additionally, while PD reaches the diffraction limit requirement on average (red star), it does exhibit approximately 100 outliers out of 1000 samples. While IS is widely spread in terms of computing time and isn't reliable, only PD follows NN performance in terms of WFE. However, the estimation time is at least 3 orders of magnitude slower compared to NN estimation. Therefore, the NN method appear to be more reliable to accurately estimate the wavefront compared to a mono-image phase diversity and image sharpening.

In this section, our primary focus has been examining NN methods and comparing them to two alternative approaches. Both phase diversity and image sharpening can naturally be improved upon. PD methods, for instance, could potentially benefit from linearized analytic phase diversity\,\cite{mocoeur2008analyse} to improve computation time and load. For IS, one could potentially investigate different image sharpness metrics that provide better performance, but also look at faster optimisation algorithms correcting one mode at a time\,\cite{debarre2008adaptive}.

\section{Towards a complete phasing sequence}\label{sec:AZI}

\begin{figure} [h!]
    \begin{center}
        \includegraphics[height = 1.85cm]{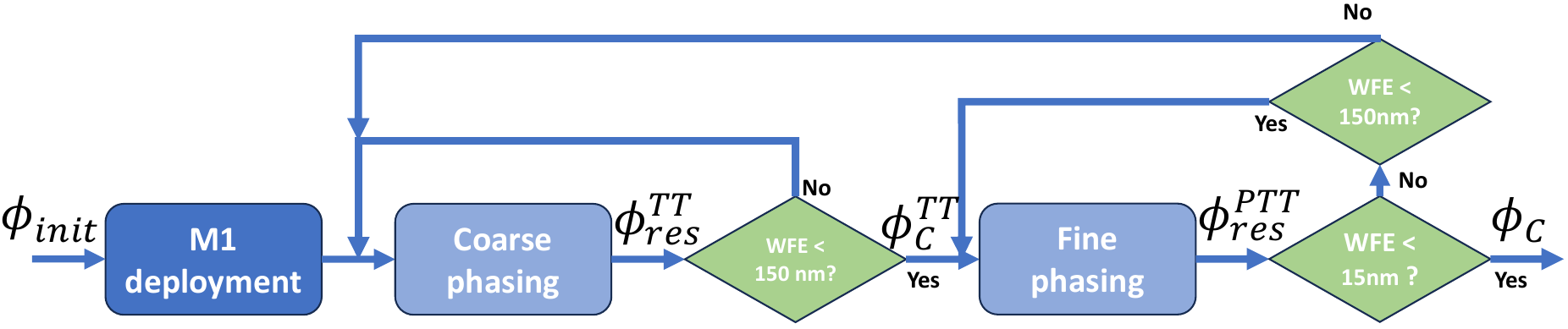}
        \caption[example] 
        {   \label{fig:phasing} 
           Diagram of the AZIMOV phasing strategy. Where $\phi_{init}$ is the initial aberrated phase,
           $\phi_{res}^{TT}$ is the residual tip-tilt phase with uncorrected piston,  $\phi_{C}^{TT}$ the phase with pre-corrected tip-tilt. $\phi_{res}^{PTT}$ the residual phase after PTT correction and $\phi_{C}$ the phase corresponding to the diffraction limit.
        }
    \end{center}
\end{figure}

Once launched, the telescope will deploy the $4$ petals with a mechanical accuracy of a few micrometers\,\cite{sauvage2020first}. The initial stage of the phasing strategy aims to mitigate the significant piston and tip-tilt errors introduced during deployment. After initial segment identification and phasing procedure recurring for instance to the ELASTIC method$\,$\cite{Schwartz, vievard2017large},the expected WFE reaches a sub-$\lambda$ amplitude over the entire pupil. We have not managed to train a suitable single NN able to retrieve high performance on large aberrations. However, we propose a solution with two NNs in cascade (a first NN for the initial large tip-tilt sensing ignoring piston, and a second NN for the piston and the remaining PTT).
The AZIMOV phasing strategy relies on two main steps. First, after deployment as the TT errors are large (as well as the piston errors), the spots do not superimpose and the piston errors do not affect (or only marginally) the resulting image. The primary mirror needs to be first corrected in tip-tilt in order to recenter each of the 4 PSFs generated by the individual segments. This allows the 4 PSFs to be superimposed on top of each other, resulting in a single PSF in the focal plane image. After this superposition, the Piston error can be estimated. Finally, the remaining piston and residual tip-tilt errors can be estimated using the NN presented in part\,\ref{sec:Res}.
Our correction can be applied iteratively in closed-loop at SNR=100, to process real-time wavefront estimation. The proposed phasing strategy can be found Fig\ref{fig:phasing}.

\begin{figure}[h!]
\begin{center}
\begin{subfigure}[b]{0.45\textwidth}
   \includegraphics[width=8cm]{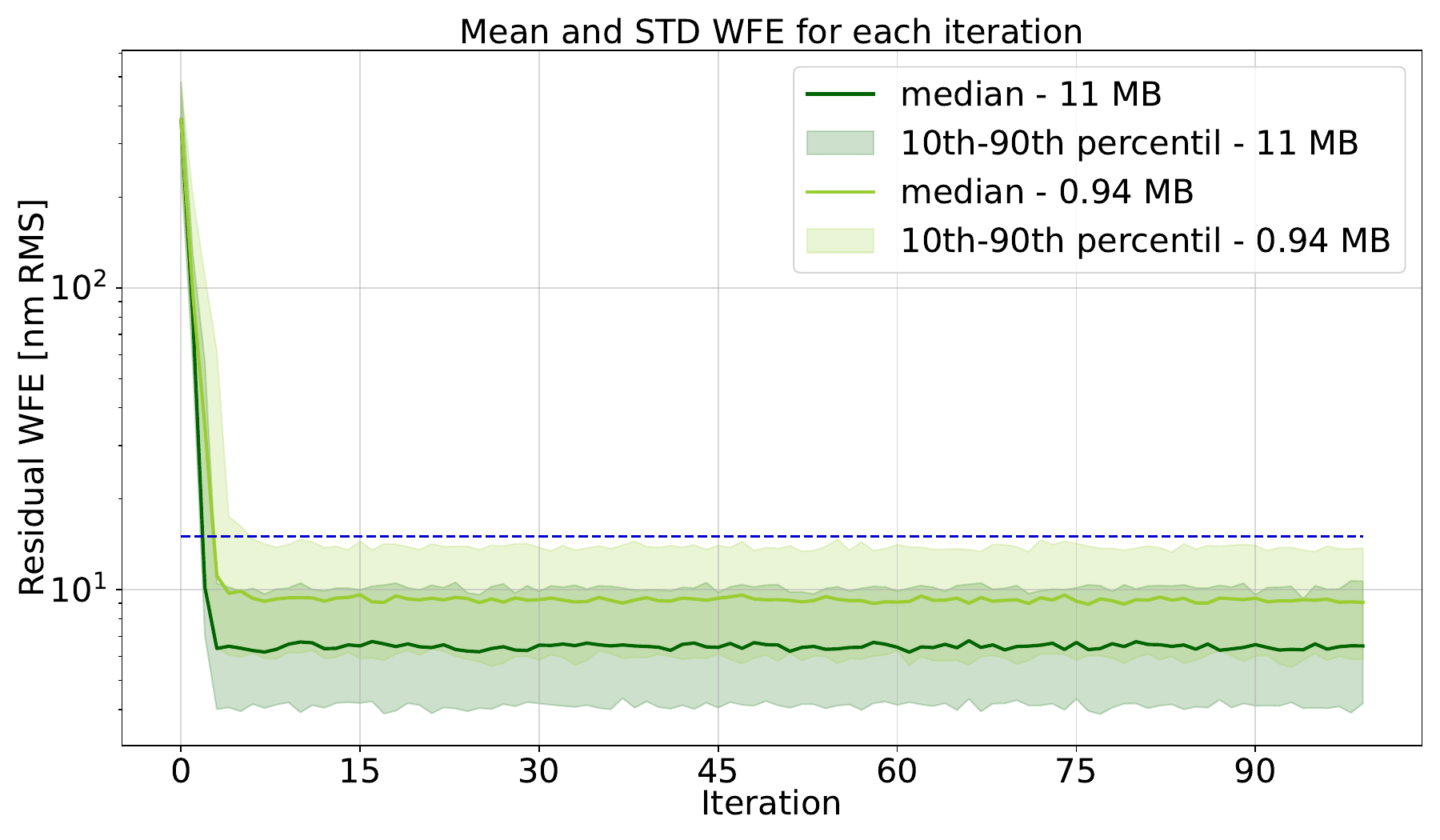}
   \caption{Average residual WFE for 100 iterations in a closed-loop over 1000 PSFs of the testset. The green shaded area represents represents the bounds of the 10th and 90th percentiles.}
   \label{fig:WFE_CL}
\end{subfigure}
\begin{subfigure}[b]{0.45\textwidth}
   \includegraphics[width=8cm]{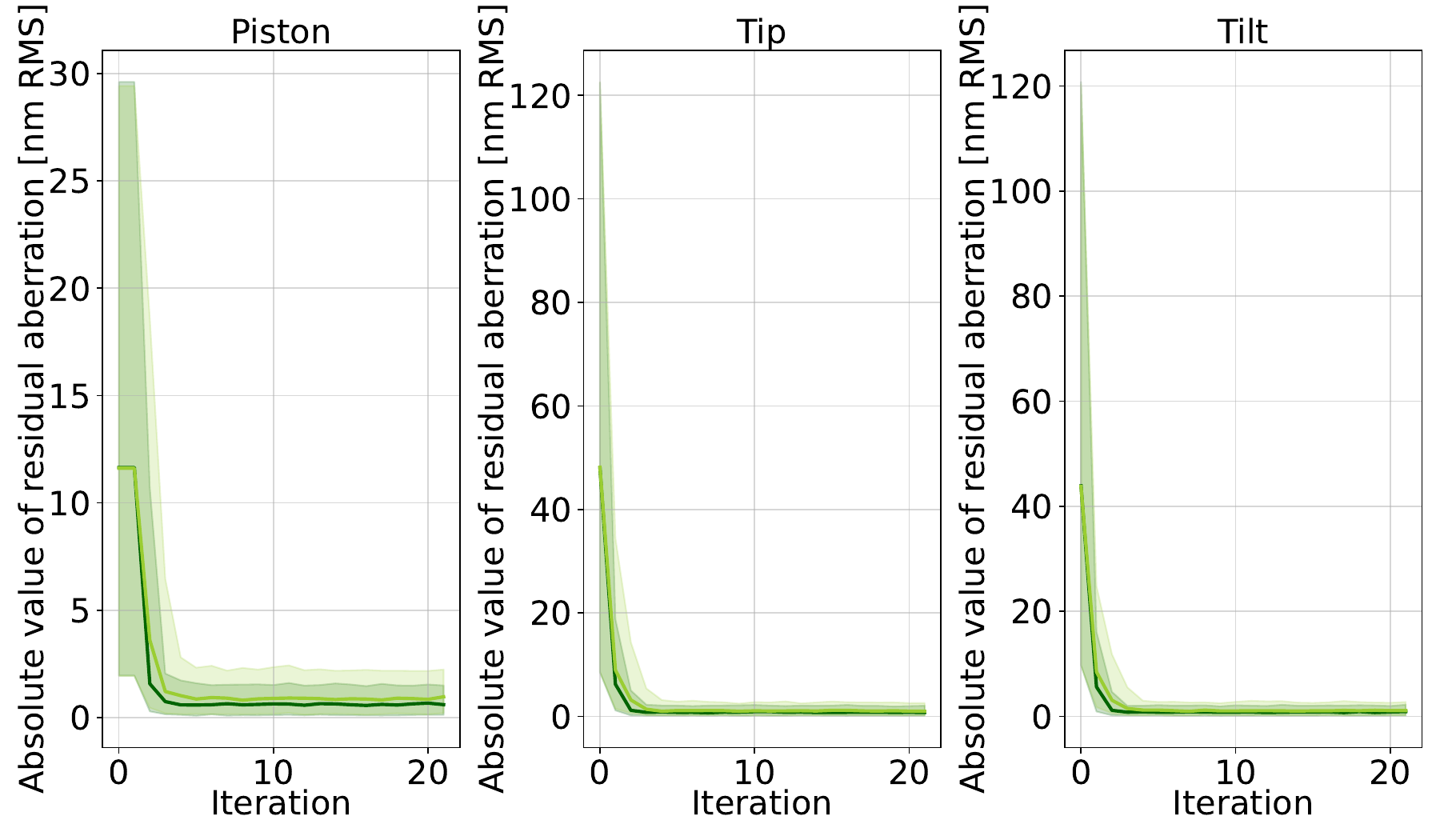}
   \caption{Average PTT residuals for 15 iterations in a closed-loop. The green shaded area represents the bounds of the 10th and 90th percentiles. Iteration 0 corresponds to the initial WFE.}
   \label{fig:coef_CL}

\end{subfigure}
\caption{Closed-loop residuals for AZIMOV deployment regime, mean and STD results over the same 1000 PSFs of the testset at SNR=100. Both figures contains the same legend.}
\end{center}
\end{figure}

The first iteration corrects for the coarse tip-tilt errors. Fig.\,\ref{fig:coef_CL} shows this while the piston errors remain unchanged. Then after the second iteration, the NN responsible for fine PTT phasing corrects for the remaining aberrations eventually converging towards about a nanometer of piston and residual tip-tilt resulting in $<10\unit{nm}$ WFE over the full pupil. The diffraction limit is reached after 2 or 3 iterations for the 11\,MB model, and slightly after 4 or 5 iterations for the 0.94MB model, Fig.\,\ref{fig:WFE_CL}. In the presence of noise at $SNR=100$ increases the standard deviation of the estimation error. However, the estimation remains well under the requirement and shows a great stability during 100 iterations for both models. In addition, it is important to notice that we limited the tip-tilt amplitude due to the amount expected in the deployment, yet larger tip-tilt could be easily added and learned by our method as long as the PSFs remain in the field of view of the detector.

\section{Conclusion}

\label{sec:Ccl}
This study explored the  performance of Neural Networks for phasing a 4-segment deployable space telescope with piston, tip and tilt aberrations in the range of ($20-120 \unit{nm\ RMS}$). The existing state-of-the-art algorithms are deemed too complex because of their iterative nature and large computational burden for space applications. Alternatively, simplified NN architectures are explored, which exhibit promising results for accurately estimating the piston-tip-tilt phasing errors. A five-convolutional layer based Resnet has been implemented \ cor{demonstrating} robustness even in the presence of additional high order aberrations and noise, by reaching the diffraction limit in different cases at SNR=50 and for 30\,nm\,RMS and 60\,nm\,RMS of HO aberrations. Additionally, it outperforms concurrent methods such as mono-image phase diversity and image sharpening in terms of computation time by at least a factor 100 with comparable noise propagation performance. Our implementation shows great performance and stability once applied in closed-loop, for both coarse and fine phasing, using a relatively simple architecture allowing for fast inference times of about few milliseconds per step.\\
However, using a single in-focus image for FPWFSing appears to limit the piston capture range between -$\frac{\lambda}{4}$ and $\frac{\lambda}{4}$. Nevertheless, this limitation might be overcome by using several input images. For instance, capturing two images at different wavelengths or alternatively using an in-focus and an out-of-focus image could overcome the lambda ambiguity and extend the capture range. Ultimately, the NN approach provides a more generic solution, both in terms of aberration capture range and potentially in terms of scene (point source vs extended).
\\
In our case of simulated data, NN are driven by the image formation model. Once on-sky data will be available, fine-tuning our method can enhance the performance to fit better to realistic cases. \\
Finally, our work presents results employing a point source. Finding such a source may be possible during the initial calibration stage. However, in the context of Earth observation, obtaining such a point source may be difficult in practice as objects of interest are typically more complex and extended. Further work will be directed towards the study of extended objects in regions of adequate contrast to convey sufficient phase information.

\section{Acknowledgments}
This work is supported by European Structural and Investment Funds in the FEDER component, through the Operational Competitiveness and Internationalization Programme (COMPETE 2020) [Project nº 047264; Funding Reference: POCI-01-0247-FEDER-047264]. \\
Authors are acknowledging the support by the Action Spécifique Haute Résolution Angulaire (ASHRA) of CNRS/INSU co-funded by CNES. This project is supported by ANR WOLF.\\
This research has made use of computing facilities operated by CeSAM data center at LAM, Marseille, France.

\section{Disclosure}
The authors declare no conflicts of interest.

\bibliography{main} 

\begin{thebibliography}{10}
\newcommand{\enquote}[1]{``#1''}

\bibitem{sabelhaus2004overview}
P.~A. Sabelhaus and J.~E. Decker, {\protect\JournalTitle{Optical, Infrared, and Millimeter Space Telescopes}} \textbf{5487}, 550 (2004).

\bibitem{Schwartz}
N.~Schwartz, W.~Brzozowski, M.~Milanova, K.~Morris, S.~Todd, Z.~Ali, J.-F. Sauvage, A.~Ward, D.~Lunney, and D.~Macleod, \enquote{High- resolution deployable cubesat prototype,} in \emph{Space Telescopes and Instrumentation 2020: Optical, In- frared, and Millimeter Wave, Dec 2020, Online Only, United States,}  (2020), p.~96.

\bibitem{born2013principles}
M.~Born and E.~Wolf, \emph{Principles of optics: electromagnetic theory of propagation, interference and diffraction of light} (Elsevier, 2013).

\bibitem{schwartz20226u}
N.~Schwartz, W.~Brzozowski, Z.~Ali, M.~Milanova, K.~Morris, C.~Bond, J.~Keogh, D.~Harvey, L.~Bissell, J.-F. Sauvage \emph{et~al.}, \enquote{6u cubesat deployable telescope for optical earth observation and astronomical optical imaging,} in \emph{Space Telescopes and Instrumentation 2022: Optical, Infrared, and Millimeter Wave,} , vol. 12180 (SPIE, 2022), pp. 1073--1084.

\bibitem{sauvage2020first}
J.-F. Sauvage, N.~Schwartz, S.~Vievard, A.~Bonnefois, M.-T. Velluet, C.~Correia, F.~Cassaing, T.~Fusco, V.~Michau, J.-C. Krapez \emph{et~al.}, \enquote{First error budget of a deployable cubesat telescope,} in \emph{Space Telescopes and Instrumentation 2020: Optical, Infrared, and Millimeter Wave,} , vol. 11443 (SPIE, 2020), pp. 554--564.

\bibitem{chanan1986segment}
G.~A. Chanan, J.~E. Nelson, and T.~S. Mast, \enquote{Segment alignment for the keck telescope primary mirror,} in \emph{Advanced Technology Optical Telescopes III,} , vol. 628 (SPIE, 1986), pp. 466--471.

\bibitem{haffert2022phasing}
S.~Y. Haffert, L.~M. Close, A.~D. Hedglen, J.~R. Males, M.~Kautz, A.~H. Bouchez, R.~Demers, F.~Quir{\'o}s-Pacheco, B.~N. Sitarski, K.~Van~Gorkom \emph{et~al.}, {\protect\JournalTitle{Journal of Astronomical Telescopes, Instruments, and Systems}} \textbf{8}, 021513 (2022).

\bibitem{acton2022phasing}
D.~S. Acton, S.~Knight, M.~Carrasquilla, N.~Weiser, M.~Masciarelli, S.~Jurczyk, G.~Rapp, J.~Mueckay, E.~Wolf, J.~Murphy \emph{et~al.}, \enquote{Phasing the webb telescope,} in \emph{Space Telescopes and Instrumentation 2022: Optical, Infrared, and Millimeter Wave,} , vol. 12180 (SPIE, 2022), pp. 303--326.

\bibitem{perrin2016preparing}
M.~D. Perrin, D.~S. Acton, C.-P. Lajoie, J.~S. Knight, M.~D. Lallo, M.~Allen, W.~Baggett, E.~Barker, T.~Comeau, E.~Coppock \emph{et~al.}, \enquote{Preparing for jwst wavefront sensing and control operations,} in \emph{Space Telescopes and Instrumentation 2016: Optical, Infrared, and Millimeter Wave,} , vol. 9904 (SPIE, 2016), pp. 142--160.

\bibitem{2017JATIS...3c9001L}
M.~P. {Lamb}, C.~{Correia}, J.-F. {Sauvage}, J.-P. {V{\'e}ran}, D.~R. {Andersen}, A.~{Vigan}, P.~L. {Wizinowich}, M.~A. {van Dam}, L.~{Mugnier}, and C.~{Bond}, {\protect\JournalTitle{Journal of Astronomical Telescopes, Instruments, and Systems}} \textbf{3}, 039001 (2017).

\bibitem{mugnier2006phase}
L.~M. Mugnier, A.~Blanc, and J.~Idier, {\protect\JournalTitle{Advances in Imaging and Electron Physics}} \textbf{141}, 1 (2006).

\bibitem{martinache2013asymmetric}
F.~Martinache, {\protect\JournalTitle{Publications of the Astronomical Society of the Pacific}} \textbf{125}, 422 (2013).

\bibitem{rossi2022machine}
F.~Rossi, C.~Plantet, A.-L. Cheffot, G.~Agapito, E.~Pinna, and S.~Esposito, \enquote{Machine learning techniques for piston sensing,} in \emph{Adaptive Optics Systems VIII,} , vol. 12185 (SPIE, 2022), pp. 1669--1674.

\bibitem{herbel2018fast}
J.~Herbel, T.~Kacprzak, A.~Amara, A.~Refregier, and A.~Lucchi, {\protect\JournalTitle{Journal of Cosmology and Astroparticle Physics}} \textbf{2018}, 054 (2018).

\bibitem{paine2018machine}
S.~W. Paine and J.~R. Fienup, {\protect\JournalTitle{Optics letters}} \textbf{43}, 1235 (2018).

\bibitem{wang2021deep}
Y.~Wang, F.~Jiang, G.~Ju, B.~Xu, Q.~An, C.~Zhang, S.~Wang, and S.~Xu, {\protect\JournalTitle{Optics Express}} \textbf{29}, 25960 (2021).

\bibitem{rajaoberison2022machine}
H.~F. Rajaoberison, J.~S. Tang, and J.~R. Fienup, \enquote{Machine learning wavefront sensing for the james webb space telescope,} in \emph{Space Telescopes and Instrumentation 2022: Optical, Infrared, and Millimeter Wave,} , vol. 12180 (SPIE, 2022), pp. 2210--2216.

\bibitem{andersen2020image}
T.~Andersen, M.~Owner-Petersen, and A.~Enmark, {\protect\JournalTitle{Journal of Astronomical Telescopes, Instruments, and Systems}} \textbf{6}, 034002 (2020).

\bibitem{orban2021focal}
G.~Orban De~Xivry, M.~Quesnel, P.~Vanberg, O.~Absil, and G.~Louppe, {\protect\JournalTitle{Monthly Notices of the Royal Astronomical Society}} \textbf{505}, 5702 (2021).

\bibitem{Nishizaki}
Y.~Nishizaki, M.~Valdivia, R.~Horisaki, K.~Kitaguchi, M.~Saito, J.~Tanida, and E.~Vera, \emph{Deep learning wavefront sensing} (Opt. Express 27, New York, 2019).

\bibitem{quesnel2022deep}
M.~Quesnel, G.~O. de~Xivry, G.~Louppe, and O.~Absil, {\protect\JournalTitle{arXiv preprint arXiv:2210.00632}}  (2022).

\bibitem{pope2014demonstration}
B.~Pope, N.~Cvetojevic, A.~Cheetham, F.~Martinache, B.~Norris, and P.~Tuthill, {\protect\JournalTitle{Monthly Notices of the Royal Astronomical Society}} \textbf{440}, 125 (2014).

\bibitem{vievard2017large}
S.~Vievard, F.~Cassaing, and L.~Mugnier, {\protect\JournalTitle{JOSA A}} \textbf{34}, 1272 (2017).

\bibitem{dohlen2011sphere}
K.~Dohlen, F.~Wildi, J.~Beuzit, P.~Puget, D.~Mouillet, A.~Baruffolo, A.~Boccaletti, J.~Charton, R.~Claudi, A.~Costille \emph{et~al.}, {\protect\JournalTitle{Adaptive Optics for Extremely Large Telescopes (AO4ELT)}} \textbf{2011} (2011).

\bibitem{goodfellow2016deep}
I.~Goodfellow, Y.~Bengio, and A.~Courville, \emph{Deep learning} (MIT press, 2016).

\bibitem{simonyan2014very}
K.~Simonyan and A.~Zisserman, {\protect\JournalTitle{arXiv preprint arXiv:1409.1556}}  (2014).

\bibitem{he2016deep}
K.~He, X.~Zhang, S.~Ren, and J.~Sun, \enquote{Deep residual learning for image recognition,} in \emph{Proceedings of the IEEE conference on computer vision and pattern recognition,}  (2016), pp. 770--778.

\bibitem{blanc2002identification}
A.~Blanc, \enquote{Identification de r{\'e}ponse impulsionnelle et restauration d'images: apport de la diversit{\'e} de phase,} Ph.D. thesis, Paris 11 (2002).

\bibitem{polyak1969conjugate}
B.~T. Polyak, {\protect\JournalTitle{USSR Computational Mathematics and Mathematical Physics}} \textbf{9}, 94 (1969).

\bibitem{powell1994direct}
M.~J. Powell, \emph{A direct search optimization method that models the objective and constraint functions by linear interpolation} (Springer, 1994).

\bibitem{mocoeur2008analyse}
I.~Mocoeur, \enquote{Analyse de front d'onde en plan focal: d{\'e}veloppement d'algorithmes temps-r{\'e}el et application au cophasage de t{\'e}lescopes multipupilles imageurs,} Ph.D. thesis, Universit{\'e} Paris Sud-Paris XI (2008).

\bibitem{debarre2008adaptive}
D.~D{\'e}barre, E.~J. Botcherby, M.~J. Booth, and T.~Wilson, {\protect\JournalTitle{Optics express}} \textbf{16}, 9290 (2008).

\end{thebibliography}

\end{document}